\documentclass[%
 reprint,
superscriptaddress,
floatfix
 amsmath,amssymb,
 aps,
 prb,
]{revtex4-1}

\usepackage{physics}
\usepackage[dvipsnames]{xcolor}

\usepackage[utf8]{inputenc}
\usepackage{etoolbox}
\apptocmd{\thebibliography}{\raggedright}{}{}

\newcommand\ee{\mathrm{e}}

\newcommand\taup{\tau^{\prime}}
\renewcommand\dd[1]{\mathrm{d}#1\:}

\newcommand\lambdap{\lambda^{\prime}}

\newcommand\dee{d^{\vphantom{\dagger}}}
\newcommand\ddagg{d^{\dagger}}
\newcommand\eff{f^{\vphantom{\dagger}}}
\newcommand\fdagg{f^{\dagger}}
\newcommand\Hc{\mathrm{H.\,c.}}

\newcommand\bigO{\mathcal{O}}
\newcommand\loc{\mathrm{loc}}
\newcommand\bath{\mathrm{bath}}

\newcommand\mc[1]{\mathcal{#1}}
\newcommand\mn[1]{\text{#1}}
\newcommand\nk[1]{{\left( #1 \right)}}
\newcommand\ek[1]{{\left[ #1 \right]}}

\usepackage{multirow}
\usepackage{graphicx}
\usepackage{dcolumn}
\usepackage{bm}
\usepackage{hyperref}
\begin{document}


\title{State- and superstate sampling in hybridization-expansion \\ continuous-time quantum Monte Carlo}

\author{Alexander Kowalski}
\affiliation{Institut f\"ur Theoretische Physik und Astrophysik, \\
Universit\"at W\"urzburg, Am Hubland, D-97074 W\"urzburg, Germany}%

\author{Andreas Hausoel}
\affiliation{Institut f\"ur Theoretische Physik und Astrophysik, \\
Universit\"at W\"urzburg, Am Hubland, D-97074 W\"urzburg, Germany}%

\author{Markus Wallerberger}
\affiliation{Department of Physics, University of Michigan, Ann Arbor, MI 48109, USA}

\author{Patrik Gunacker}
\affiliation{Institute of Solid State Physics, 
   TU Wien, 1040 Vienna, Austria }%

\author{Giorgio Sangiovanni}
\affiliation{Institut f\"ur Theoretische Physik und Astrophysik, \\
Universit\"at W\"urzburg, Am Hubland, D-97074 W\"urzburg, Germany}%


\begin{abstract}

Due to the intrinsic complexity of the quantum many-body problem, quantum Monte Carlo algorithms and their corresponding Monte Carlo configurations can be defined in various ways.
Configurations corresponding to few Feynman diagrams often lead to severe sign problems.
On the other hand, computing the configuration weight becomes numerically expensive in the opposite limit in which many diagrams are grouped together.
Here we show that for continuous-time quantum Monte Carlo in the hybridization expansion the efficiency can be substantially improved by dividing the local impurity trace into fragments, which are then sampled individually.
For this technique, which also turns out to preserve the fermionic sign, a modified update strategy is introduced in order to ensure ergodicity.
Our (super)state sampling is particularly beneficial to calculations with many $d$-orbitals and general local interactions, such as full Coulomb interaction.
For illustration, we reconsider the simple albeit well-known case of a degenerate three-orbital model at low temperatures. This allows us to quantify the coherence properties of the ``spin-freezing'' crossover, even close to the Mott transition.
\end{abstract}

\maketitle


\section{Introduction}

Continuous-time quantum Monte Carlo algorithms are state-of-the-art, numerically exact methods for the solution of the Anderson impurity model (AIM) \cite{RubtsovPRB,werner-continuous-time-2006,haule-quantum-2007,gull-continuous-time-2011}.
These are widely used for the description of the physics of magnetic impurities, Kondo systems, transport through quantum junctions and are also employed as auxiliary models in dynamical mean field theory (DMFT) calculations for lattice models of correlated electron systems. 
Several high-level open source implementations of DMFT and of its merger with density functional theory have been recently made available \cite{shinaoka-continuous-time-2017, triqs, haule_code, iqist, w2dynamics}.

One of the most successful flavors of continuous-time quantum Monte Carlo algorithms is the strong-coupling hybridization expansion (CT-HYB)\cite{wernerPRB74}.  CT-HYB is the method of choice for multiorbital impurity models with general interactions because one observes only a moderate sign problem provided that the bath problem has sufficient symmetry.  This is because CT-HYB splits each Monte Carlo configuration into a noninteracting bath part and a fully interacting impurity part, and solves the impurity part using an exact diagonalization/FullCI-type method.  
However, the dimension of the impurity Hamiltonian grows exponentially with the number of orbitals, and so does the computational effort with it. In practice, correlated $d$- or $f$-shells as well as small correlated molecules can be treated with CT-HYB.

Yet, reaching low temperatures is still challenging due to the quadratic scaling of the impurity problem with inverse temperature. This follows from the fact that the mean order of diagrammatic expansion grows linearly with inverse temperature, and both the computational cost of evaluating a single configuration as well as the observed autocorrelation time between configurations scale linearly with expansion order. While the exponential scaling with the number of orbitals and the quadratic scaling with the inverse temperature are intrinsic to the local problem, potentially model-dependent improvements to the prefactor of this overall scaling can be achieved.

Common approaches to such optimization are block diagonalization of
the local Hamiltonian using conserved quantities~\cite{haule-quantum-2007,parragh-conserved-2012} and binning, tree\cite{gull-continuous-time-2011}, or equivalent\cite{haule-lazy-trace} algorithms in so-called ``matrix-matrix'' implementations of CT-HYB.
Additionally, with a similar motivation as for our method, outer truncation of the local trace to the few dominant contributions and calculation of those with more efficient sparse-matrix methods has been applied particularly to large systems at low temperatures\cite{werner-krylov-2009}. 
Other more advanced strategies are local updates in imaginary time \cite{shinaoka-hybridization-2014}, 
a fast-rejection/acceptance algorithm by calculating upper/lower boundaries of the weight 
\cite{haule-lazy-trace}, or a partial summation of diagrams to extract more information out of one 
Monte Carlo configuration \cite{Augustinsky2013}.

Here we consider a matrix-vector version of the CT-HYB algorithm as implemented in the \emph{w2dynamics} package \cite{w2dynamics} and investigate the possibility of sampling the sum over the eigenstates of the local impurity in the Monte Carlo simulation.
A hard outer truncation of high energy states (so far typically used in calculations with \emph{w2dynamics}) constitutes an approximation and it is unclear whether it retains ergodicity.
The approach proposed here, instead, is numerically exact and furthermore exceeds the performance benefits of hard truncation substantially.

We formulate two versions: the ``superstate''-sampling algorithm, where states grouped by the blocks of the Hamiltonian
are sampled together, and the ``state''-sampling algorithm, where each many-body state of the impurity is sampled
individually. Conceptually, these methods can be interpreted as an equivalent of the segment-algorithm for general interactions.
Our improvements touch the core of the exponential scaling of CT-HYB and manage to significantly reduce the computation time of a Monte Carlo weight. 
Furthermore, they are in principle compatible with all of the other above-mentioned algorithmic improvements. 
Using a five-orbital AIM with the most general form of the electron-electron Coulomb interaction as an example, we achieve speed-up factors verging on three orders of magnitude. 

First, we review the basic formulas of CT-HYB in Sec.~\ref{s_cthyb}.
In Secs.~\ref{s_superstatessampling} and \ref{sec:statesampling}, the superstate and state sampling
methods are introduced. 
In Sec.~\ref{sbenchmarks}, we comment on the performance and the average sign, while
in Sec.~\ref{asdfspin} we demonstrate the capabilities of the (super)state algorithm
with a simple physical example.

\section{Hybridization expansion}
\label{s_cthyb}

Let us start with a brief review of the hybridization-expansion continuous-time quantum Monte Carlo algorithm, focusing on the formulas needed to explain our changes. For a complete introduction to this method, see Ref.~\onlinecite{gull-continuous-time-2011}.
We are concerned with the solution of a multiorbital Anderson impurity model, whose Hamiltonian can be written as
\begin{equation}
H = H_\loc[\dee, \ddagg] + \sum_{p\lambda} (V_{p\lambda} \fdagg_p \dee_\lambda + \Hc)
+ \sum_p \tilde E_p \fdagg_p \eff_p,
\end{equation}
where $\dee_\lambda$ annihilates a fermion on the impurity, which
consists of spin orbitals $\lambda\in\{1,\ldots,2N_\text{orb}\}$, 
and $\eff_p$ annihilates
a fermion on the bath, where the quantum number $p$ can be continuous. $V_{p\lambda}$
and $\tilde E_p$ parametrize the hybridization and bath levels, respectively,
while $H_\loc$ is a generic interacting local (impurity) Hamiltonian.

The expansion of the partition function $Z = \Tr\ee^{-\beta H}$ in terms of
the bath hybridization can be written as:\cite{gull-continuous-time-2011}
\begin{equation}\begin{split}
&Z = Z_0 \sum_{k=0}^\infty
  \sum_{\lambda_1,\lambdap_1} \int_0^\beta\!\dd{\tau_1} \int_0^\beta\!\dd{\taup_1}
  \cdots\sum_{\lambda_k,\lambdap_k} \int_{\tau_{k-1}}^\beta\!\!\!\dd{\tau_k}
  \int_{\taup_{k-1}}^\beta\!\!\!\!\dd{\taup_k}\\
&\!\times \Tr\!\left[ T_\tau \ee^{-\beta H_\loc} \prod_{i=1}^k \ddagg_{\lambda_i}\!(\tau_i)\ \dee_{\lambdap_i}\!(\taup_i) \right]
   \det[\Delta_{\lambda_i\lambdap_j}\!(\tau_i - \taup_j)]_{ij},
\end{split}
\label{zexp}
\end{equation}
where $Z_0$ is the partition
function for the bath part,
$\beta = 1/T$ is the inverse temperature, $T_\tau$ denotes path ordering
in imaginary time, and $\Delta(\tau) = V^\dagger (\partial_\tau - \tilde E)^{-1} V$
is the hybridization function, which encodes the total retardation effect
of the bath on the local fermions.

Schematically, Eq.~(\ref{zexp}) can be written as follows:
\begin{equation}
   Z=\sum_{\mc{C}} w_{\mn{loc}}(\mc{C}) w_{\mn{bath}}(\mc{C}),
   \label{zsym}
\end{equation}
where
$\mc C := (\tau_1, \taup_1, \lambda_1, \lambdap_1, \ldots, \tau_k, \taup_k, \lambda_k, \lambdap_k)$
denotes an infinitesimal term in the expansion, i.e.,
\begin{equation}
\!\sum_{\mc{C}} := \sum_{k=0}^\infty
  \sum_{\lambda_1,\lambdap_1} \int_0^\beta\!\dd{\tau_1} \int_0^\beta\!\dd{\taup_1}
  \!\cdots\! \sum_{\lambda_k,\lambdap_k} \int_{\tau_{k-1}}^\beta\!\!\!\dd{\tau_k}
  \int_{\taup_{k-1}}^\beta\!\!\!\!\dd{\taup_k}.
  \label{csum}
\end{equation}
In the conventional continuous-time hybridization expansion quantum Monte Carlo
(CT-HYB) algorithm, each $\mc C$ is taken as a Monte Carlo configuration, and
the sum (\ref{csum}) is performed using Markov chain Monte Carlo:
\begin{equation}
   Z=\underbrace{\sum_{\mc{C}}}_{\mn{QMC}} w_{\mn{loc}}(\mc{C}) w_{\mn{bath}}(\mc{C}).
\end{equation}

With $w_\bath = \det[\ldots]$ we denote the bath part, corresponding to a determinant of
noninteracting hybridization functions, which can be computed in $\bigO(k^3)$ and
updated in $\bigO(k^2)$ time. The quantity
\begin{align}
   w_{\mn{loc}}(\mathcal{C}) &= \sum_{s} \bra{s}{ \hat{\mathcal{C}} }\ket{s} \label{g_loc_trace2} \\
   &:= \sum_s \Big\langle s \Big| T_\tau \mn{e}^{-\beta H_{\mn{loc}}} \prod_{i=1}^k  d_{\lambda_i}^\dagger(\tau_i) d_{\lambda_i'}(\tau_i')
   \Big| s \Big\rangle
   \label{g_loc_trace}
\end{align}
is the local weight or local trace of a configuration, where $s$ indexes the
$4^{N_\text{orb}}$ many-body eigenstates of the local impurity Hamiltonian $H_\loc$, and
we have introduced the shorthand $\hat{\mc C}$ for the sequence of local
operators of the current configuration. A na\"ive implementation of Eq.~(\ref{g_loc_trace}) involves $k$ multiplications
of $4^{N_\text{orb}} \times 4^{N_\text{orb}}$ matrices, which scales as $\bigO(k \exp(\alpha N_\text{orb}))$ with a constant $\alpha$.  
Reducing the computational impact of the
calculation of $w_{\mn{loc}}(\mc{C})$ is thus usually the main objective of
optimizing CT-HYB codes.

In general, the local Hamiltonian $H_\loc$ conserves a set of quantum numbers.
Consequently, the many-body Hilbert space can be partitioned into a set of linear subspaces  $\{\mathcal S\}$,
so-called ``superstates'',\cite{haule-quantum-2007} and the Hamiltonian can be
brought into a block-diagonal form with respect to these superstates.
We can write Eq.~(\ref{g_loc_trace2}) as:
\begin{equation}
 w_{\mn{loc}}(\mathcal{C}) = \sum_{\mathcal S} w_{\loc,\mathcal S} (\mathcal{C})
 := \sum_{\mathcal S} \sum_{s\in \mathcal S} \bra{s}{ \hat{\mathcal{C}} }\ket{s}.
 \label{wlocS}
\end{equation}
In defining the quantum numbers, we impose the additional requirement that the impurity operators do not take a state from one superstate to more
than one other superstate, thereby possibly merging multiple blocks of the Hamiltonian into one superstate. 
This implies that $w_{\loc,\mathcal S}$ can be calculated independently for each $\mathcal S$.  
The scaling is now controlled by the size of the largest superstate (in the worst case), i.e., by something much smaller than $4^{N_\text{orb}}$.
Since the application of an impurity operator corresponds to a
one-to-one mapping between different superstates (and giving zero if it
violates the Pauli principle), a further optimization is possible: For
each superstate $\mathcal{S}$, one can follow the sequence of
superstates by using this mapping starting with $\mathcal{S}$ at
$\tau = 0$ until reaching $\tau = \beta$. If one reaches zero at any
point or ends up in a different superstate at the other end,
$w_{\loc,\mathcal S}$ is exactly zero and does not need to be
calculated using possibly much more costly linear algebra. We will
refer to this procedure as \textit{quantum number checking} in the
following sections.

This concludes our overview of what we refer to as conventional CT-HYB method.

\section{Superstate sampling}
\label{s_superstatessampling}
\subsection{General description}

The main idea of this work is to transform the deterministic summation over
the eigenstates of the impurity in Eq.~(\ref{g_loc_trace2})
into a stochastic summation.  The original Monte Carlo configuration
is split into many ``smaller'' weights.  While having been proposed\cite{wallerberger-w2dynamics-2016},
it has never been implemented to the best of our knowledge.

We focus on two strategies in particular, one that partitions the sum
into subsets by quantum numbers and one that breaks it up
entirely. 
We call them

\begin{enumerate}
 \item ``Superstate sampling'': the summation over all superstates $\mc{S}$ is
       now done by Monte Carlo sampling:
\begin{equation}
   Z=\underbrace{\sum_{\mc{C}} \sum_\mc{S}}_{\mn{QMC}} \sum_{s\in\mathcal S}
   \bra{s}{ \hat{\mathcal{C}} }\ket{s} w_{\mn{bath}}(\mc{C}),
\end{equation}
where each Monte Carlo configuration now contains the trace over all states $s$
within a superstate.

 \item ``State sampling'': the summation over all states $s$ in Eq.~(\ref{wlocS})
       is now done by Monte Carlo sampling:
\begin{equation}
   Z=\underbrace{\sum_{\mc{C}} \sum_{\mc{S}} \sum_{s\in \mc{S}}}_{\mn{QMC}}
   \bra{s}{ \hat{\mathcal{C}} }\ket{s} w_{\mn{bath}}(\mc{C}),
\end{equation}
where each Monte Carlo configuration now contains the local configuration
evaluated for a single outer state $s$.
\end{enumerate}
In this section, we will focus on superstate sampling, while state sampling
will be discussed in Sec.~\ref{sec:statesampling}.

The fragmentation of the sum reduces the amount of calculations needed for one local weight and allows us to move faster through phase space.
It is thus particularly beneficial in systems with low symmetry, which can have many superstates with small but nonzero contributions to the local weight.
On the other hand, if many quantum numbers can be used in a calculation with the conventional sampling, an increase in $\beta$ results in an effective reduction of the number of possible outer superstates. This ``help'' is a side effect of the large number of operators in the trace present in the low-$T$ limit. It results in a very high chance of quantum number violation and it hence substantially restricts the room for maneuver for the outer superstates. 
The advantage of superstate sampling is therefore twofold: at any temperature an easy and natural selection of the most important outer superstates and much less need for quantum number checking, particularly beneficial at low $T$ \footnote{One may somewhat reduce the residual advantage of superstate sampling visible at large $\beta$ in Fig.~\ref{fig:speedup} upon optimizing the quantum number checking or using more sophisticated schemes, such as those proposed in Ref.~\cite{haule-lazy-trace}}.

By sampling superstates, we are sampling
a sum of terms with potentially different signs. This may induce a sign problem, which would in general be
expected to worsen exponentially with decreasing temperature.
(This is why it is important to combine all possible bath configurations into a bath determinant in CT-HYB.\cite{werner-continuous-time-2006})
Yet, we do not observe any worsening of the average sign in superstate sampling
compared to the original algorithm (cf.~Sec.~\ref{sbenchmarks}).
A heuristic argument for this can be summarized thusly:
Since the mean expansion order grows linearly with the inverse temperature $\beta$ \cite{werner-continuous-time-2006},
the average number of superstates that violate the Pauli principle increases, until at a certain $\beta$, we are often left with only one outer superstate.
For example, we have observed that for a typical metallic systems, a temperature of the order of $10^{-2}$ of the electronic bandwidth is about the point where many configurations have only
a single superstate contributing to the trace.
At such low temperatures, the local weight in conventional sampling is the sum over only one outer superstate, so switching to superstate sampling should not affect the sign.

\begin{figure}
  \centering
  \includegraphics[width=\linewidth]{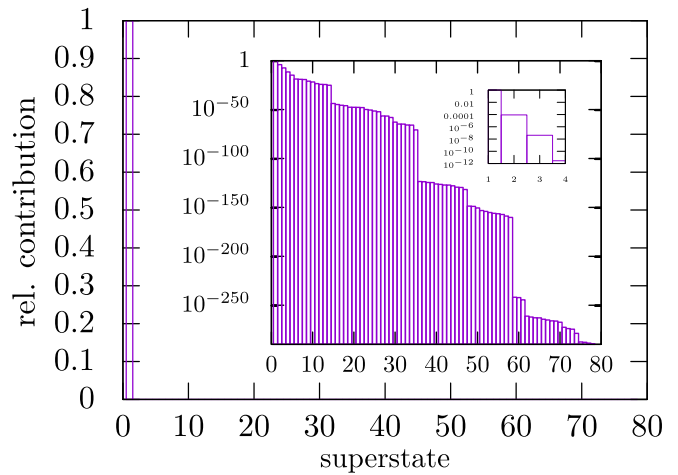}
  \caption{Average relative contributions of outer superstates to the local weight per configuration for a typical simulation with five orbitals and Kanamori interaction \cite{CNoteModel}. For comparability, the superstates are ordered by their contribution (i.e., absolute value of the part of the local weight sum from all states contained in the superstate) for each individual configuration, i.e., superstate ``1'' does not denote one specific constant superstate, but always refers to the biggest contributor.}
  \label{fig:sstweightcontrib}
\end{figure}

When multiple superstates contribute, such as in simulations at high temperatures, the switch to superstate sampling could in principle cause a difference depending on the superstates' relative weight and relative sign. To illustrate the typical superstate weight distribution in such cases, we consider relatively high-temperature simulations. For a five-orbital model with Kanamori interaction \cite{CNoteModel}, we show the average distribution of the local weight of a configuration onto the outer superstates in Fig.~\ref{fig:sstweightcontrib}. It is clear how the local weight of each configuration is strongly dominated by the contribution of one superstate. Similar results can be obtained for a simpler two-orbital model. We can therefore expect the method to be useful for high temperatures as well, as it allows us to sample configurations with their ``ideal'' outer superstates.

\subsection{Sampling and ergodicity}
\label{sec:sstoldmoves} 

\begin{figure}
  \centering
  \includegraphics{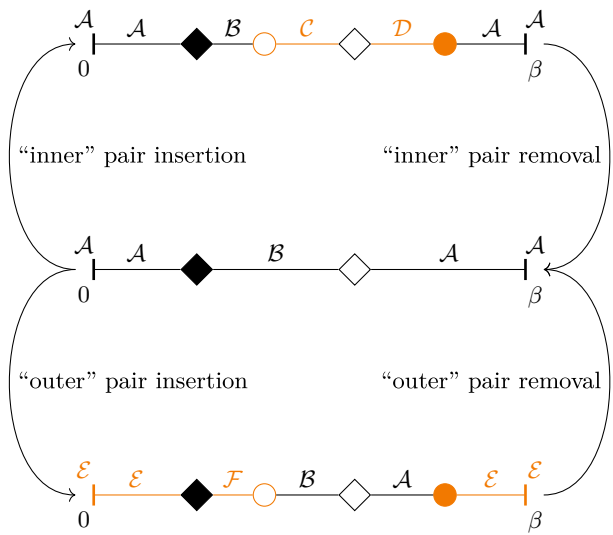}
  \caption{Three configurations represented as line segments
    with calligraphic letters denoting the superstates. 
    An ``inner'' (``outer'') pair move is shown in the upper (lower) part of the figure.
    The impurity operators are represented
    by symbols, whose filledness tells whether they are creators or
    annihilators. The imaginary times of the operators is given by
    their position along the segment, which represents the
    imaginary time axis. The letter above the $\tau\!=\!0$ mark is the outer superstate
    of the configuration.}
 \label{fig:innerouter}
\end{figure}

Since we extend the configurations in the superstate sampling method by an outer superstate, the simulation must be able to reach every configuration with nonzero weight independent of its outer superstate to preserve ergodicity.

\bigskip
\paragraph{Inner pair moves.}
While the possibility to change the outer superstate needs to be available for ergodicity, we observe that it only needs to be done comparatively rarely in the simulation. Therefore, we do not change the outer superstate when inserting or removing a pair of operators, which are the most common moves proposed. We call this variant the ``inner'' pair moves: Because we fix the outer superstate, only states that lie between the inserted or removed operators in imaginary time can change and need to be recalculated.

Consider the example of the ``inner'' pair insertion move shown in the upper half in Fig.~\ref{fig:innerouter}: The old configuration $\mc{C}_{\mn{old}}$ is the one shown in the middle panel. We want to perform an ``inner'' pair move that inserts the two orange operators with random times and flavors in the top diagram, which represents the resulting configuration $\mc{C}_{\mn{inner}}$. In this most commonly proposed type of move, new configurations $\mc{C}_{\mn{inner}}$  are only proposed with the same outer superstate as the old configuration $\mc{C}_{\mn{old}}$. 
The superstate sequence between the two new orange operators is new whereas the part outside them remains as in the old configuration.
These moves are in a sense most closely connected to the pair moves of conventional sampling as the outer superstates with the biggest contributions are not likely to change in local moves.

\bigskip

\paragraph{Outer pair moves.}
In the other local pair moves we consider, the ``outer'' moves, the prescription for the superstate sequence is the opposite. The superstates between the inserted operators are to be left unchanged, and the sequence must be continued from there to $\tau \!=\! 0$ to determine the superstate that should be used as new outer superstate. An example for an outer insertion is the move from the configuration $\mc{C}_{\mn{old}}$ in the middle of Fig.~\ref{fig:innerouter} to $\mc{C}_{\mn{outer}}$ on the bottom, where the inserted operators are the same as in the inner move example for easy comparison. Since we fix the inner part of the superstate sequence, the entire part ``outside'' of the orange operators changes.

The acceptance of outer pair moves is in practice however significantly smaller than that of the ``inner'' pair moves. This can be understood thinking about the limit of local-in-$\tau$ moves: These, in order to be considered local in the ``outer'' case, are subject to the additional constraint of having the two operators at opposite ends of the trace. In the next subsection, we discuss a more efficient way to ensure ergodicity with respect to the outer superstate, the so called global $\tau$-shift move. We will also show in Appendix~\ref{s_proof_of_ergodicity_tau_shifts} that the global $\tau$-shift moves induce an equivalence between inner and outer moves, which however does not imply equal acceptance rates.

\bigskip

A final noteworthy detail of this sampling procedure is the choice of the outer superstate for
the initial configuration at the beginning of the
simulation. While it should not influence the simulation
after thermalization, for many highly excited outer superstates the local weight is close to zero.
We thus 
select the initial outer superstates randomly with probabilities
proportional to their local weights.

Let us note that one could think of simpler techniques than the presented moves to ensure ergodicity, e.g., the addition of a move that changes \textit{only} the outer superstate, or the possibility to change the outer superstate randomly during each move. Both of these turn out to be inefficient ways compared to those we present here. Adding such a move is, however, necessary for ergodicity in a simulation of a system in the atomic limit, i.e., without hybridization, because outer (and inner) insertions would always be rejected, operators for outer removals are not present, and the move presented in the following section does not change a configuration without operators at all. Therefore, we do occasionally propose a change of just the outer superstate in configurations without any operators.

\subsection{Global $\tau$-shift moves}
\label{sec:ssttaushift}

\begin{figure}
  \centering
  \includegraphics{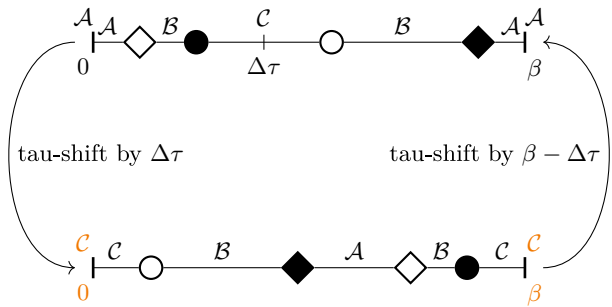}
  \caption{Diagrams representing two configurations that can be
    obtained from one another using a $\tau$-shift move. The symbols on the imaginary
    time segments represent the impurity operators, with shapes standing in for
    flavors and filledness for whether they are creators or
    annihilators. Calligraphic letters denote superstates
    with the ones above the $\tau \! = \! 0$ and $\tau \! = \! \beta$ mark giving
    the outer superstate of the configuration. When a $\tau$-shift move is
    performed, all operators are shifted according to $\tau \to \tau - \Delta \tau$ (wrapping around to $\tau \to \tau+\beta-\Delta\tau$ if necessary). \label{fig:taushift}}
\end{figure}

A global ``$\tau$-shift move'' shifts the positions of all operators in imaginary time
by a random $\Delta \tau \in \left[0,\beta\right]$, which can equivalently be thought of as a shift of the imaginary time axis. At the same time, the new outer superstate is by construction chosen to be consistent with this shift of the origin of the imaginary time axis
(see Fig. \ref{fig:taushift}).

Using just inner pair moves and the global $\tau$-shift move, a superstate sampling simulation is ergodic if and only if it is ergodic using conventional sampling with pair moves. This is because if any configuration can be built up using pair moves in conventional sampling, any configuration can be built up using inner pair moves with the outer superstate being one of the contributing ones in superstate sampling.
A proof of ergodicity can be found in Appendix~\ref{s_proof_of_ergodicity_tau_shifts}.

Let us now consider the properties of the $\tau$-shift move. Over the course of an entire simulation, the proposal probability for a specific outer superstate in this kind of move is proportional to the average relative amount of imaginary time it covers.
Since the superstate
sequence is cyclic and effectively also just shifted along the
$\tau$-axis, there is no need to perform quantum number
checking.
A global move similar to our $\tau$-shift was introduced by Shinaoka {\it et al.}~\cite{shinaoka-continuous-time-2017}
for a different technical reason. 

The proposal probabilities of a $\tau$-shift move and its reverse
are equal. The acceptance probability of this move is 1.
In Appendix~\ref{s_bath_proof} we prove that the bath determinant remains unchanged,
as the action of the $\tau$-shift on the hybridization matrix move
effectively corresponds to a number of permutations and multiplications of
rows and columns. Additionally in Appendix~\ref{s_trace_proof} there is a proof that the local
trace remains unchanged under a
combined cyclic permutation of the operators and corresponding change of the outer
superstate.

Let us discuss why $\tau$-shift moves allow us to preserve ergodicity.
Inner moves alone cannot change the outer superstate, but only
superstates in parts not including $\tau = 0 = \beta$.
$\tau$-shifts indeed move ``the section with the outer superstate'' away from 
$\tau = 0 = \beta$ and can hence shift the
operator/superstate sequence in such a way
that the outer superstate changes while
the configuration remains otherwise equivalent. In combination with
the ordinary inner pair moves it can hence change the superstate of any
section without need for outer pair moves.

Due to its favorable characteristics compared to the outer moves, we usually add just $\tau$-shift moves to the always necessary inner moves to ensure ergodicity of simulations with respect to the outer superstate. We choose to propose $\tau$-shift moves as $0.5 \%$ of all moves by default, which was also the ratio used in all calculations shown later. While a smaller ratio might improve performance, the potential speed up in usual cases would be small as the $\tau$-shift moves usually do not take up the majority of the time.

Additionally, we also allow random changes of the outer superstate in our implementation during other global moves that can be used in CT-HYB but which we do not further discuss in this paper. For such global moves as the flavor permutations used in \textit{w2dynamics}, procedures mapping the old superstate to a new proposal based on the specific move could actually be thought of, but since these moves only serve to go between badly connected areas of phase space, we prefer not to restrict them more than necessary.

\section{State sampling}
\label{sec:statesampling}
\subsection{General description}
\begin{figure}
  \centering
  \includegraphics[width=\linewidth]{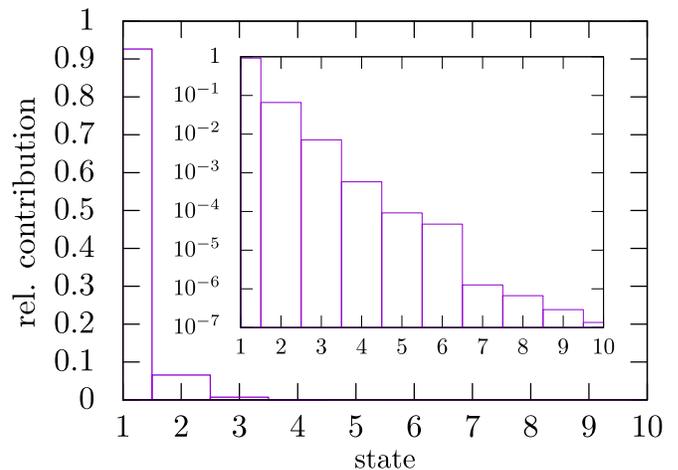}
  \caption{Average relative contribution of outer states within one superstate\cite{CNote1}, ordered by the size of contribution per configuration, normalized to the total weight of that superstate. This graph was obtained from a typical simulation of a five-orbital system with Kanamori interaction \cite{CNoteModel}.}
  \label{fig:stweightcontrib}
\end{figure}

The superstate sampling method of the last section
already significantly reduces the cost of computing a local weight.
Similarly to the predominance of a single superstate in the local weight
(Fig.~\ref{fig:sstweightcontrib}), we often find that within one superstate,
the individual eigenstates show a similar trend
(Fig.~\ref{fig:stweightcontrib}, using the same model\cite{CNoteModel}\nocite{bahlke-interplay-2018}):
The contribution of one superstate $\mathcal S$ is dominated by the
contribution of one or a small group of its eigenstates $s\in\mathcal S$.
This suggests trying to apply the principle of superstate sampling
one level deeper in the form of an ``(eigen)state sampling''.

We split the local weight
further into even smaller parts, where the summation over all states $s$ 
within a superstate $\mc{S}$ is also done as a Monte Carlo sum:
\begin{equation}
   Z=\underbrace{\sum_{\mc{C}} \sum_{\mc{S}} \sum_{s\in \mc{S}}}_{\mn{QMC}} w_{\mn{loc},s}(\mc{C}) w_{\mn{bath}}(\mc{C}),
\end{equation}
with
\begin{equation}
   w_{\mn{loc},s}(\mc{C})= \bra{s} \hat{\mc{C}} \ket{s}.
\end{equation}

To avoid confusion, let us stress at this stage that our method does \emph{not} make any assumption on the contribution of the superstates to the local weight. It is an exact sampling with no approximation involved. 

Unlike for superstate sampling, a heuristic argument from conventional sampling for a sign close to 1 cannot be given (unless a system with superstates containing a single state each is considered). This stems from the fact that at least quantitatively, the weight in state sampling is always different from the one in conventional sampling, as the latter sums up contributions from at least one entire superstate. However, as stated earlier, we empirically find the contribution of a superstate to the local weight is often dominated by one of its eigenstates. This suggests that the sign should often not be much worse than in superstate sampling, as those dominating states should be sampled considerably more often than other ones. Because the local weight in state sampling when such a dominating state $s$ is the outer state has the same sign as the local weight of the corresponding configuration with outer superstate $\mc{S} \ni s$ in superstate sampling, the maximum deterioration of the sign as compared to superstate sampling should be related to the extent to which individual states dominate the superstate weight contributions. As the time evolution further suppresses states of higher energy at lower temperature, causing the lower energy states to dominate the superstates' contributions more and more, the ``sign gap'' between the methods should also decrease with decreasing temperature.

Finally, let us note that this sampling technique only improves performance over superstate sampling if the
summation over the outer states is actually performed as the outermost summation in the numerical
implementation as well, as opposed to multiplying the operator matrices
first. While performing the summation over outer \textit{superstates} first is
common in implementations of conventional CT-HYB sampling as this allows the use of quantum
number checking for performance improvement, the summation over the
outer \textit{states} is instead often performed only after the multiplication of
the operator matrices (per superstate), as this allows the use of
optimizations employing tree structures. 
As opposed to these matrix-matrix implementations, there are so-called ``matrix-vector'' ones that
perform the summation over the outer states as the outermost one. For more detailed information, see Appendix~\ref{sec:mmmv}.

\subsection{Choice of the outer state within a superstate}
\label{sec:statesamplingdetails}

\begin{figure*}[th!]
   \centering
\includegraphics[width=\linewidth]{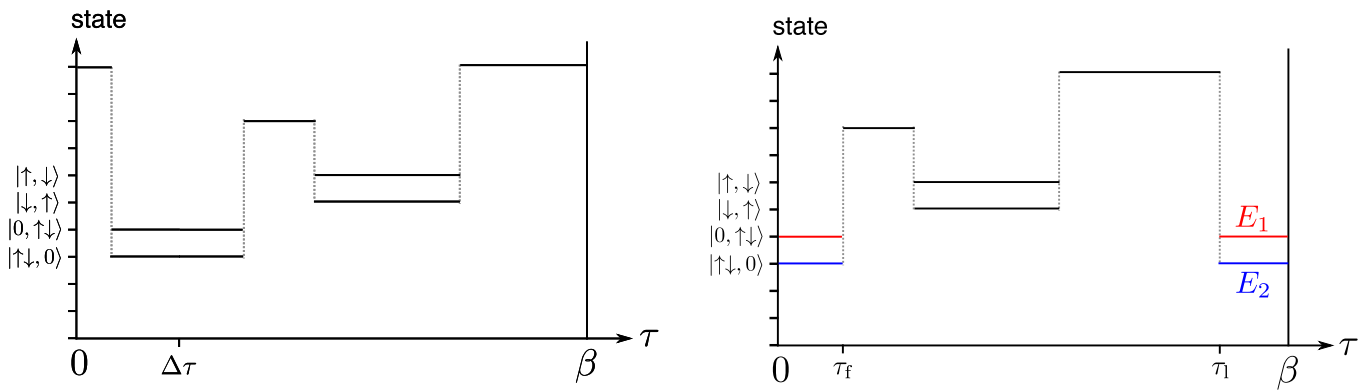}
   \caption{Left: a superstate sampling configuration for a two orbital model with Kanamori interaction.
   Bold horizontal lines denote a time-evolution of an eigenstate, the dotted vertical lines the operators 
   that cause transitions between superstates.
   Here all superstates have size 1, except the spin-flip and pair-hopping one.
Right: the configuration resulting from application of a global $\tau$-shift by $\Delta \tau$ to all operators of the one shown on the left. $\tau_f$ and $\tau_l$ denote the imaginary times of the first and last operator after the shift.
In the superstate sampling algorithm the local weight is the sum of the red and blue state as outer state of the trace,
whereas in the state sampling only one of the two is selected.\label{g_config}}
\end{figure*}

The crucial point of moves in state sampling compared to those in superstate or conventional sampling is how to choose the outer state. This is also the only point in which our state sampling moves differ from the ones we propose for superstate sampling. In superstate sampling
the choice of the new outer superstate is clear
from the way in which the superstate sequence is changed by the move.
This is the case for both outer moves and the $\tau$-shift move (cf.\ Secs.~\ref{sec:sstoldmoves} and \ref{sec:ssttaushift}). In state sampling, the situation is different because a qualitative equivalent of the superstate sequence for states does not exist.
At $\tau=0$ there is indeed an eigenstate of the Hamiltonian,
but it will change into a linear combination of multiple eigenstates after the first application of an impurity operator.

In the inner pair moves (cf.\ Sec.~\ref{sec:sstoldmoves}, Fig.~\ref{fig:innerouter}), we simply keep the outer state fixed, just as we kept the outer superstate fixed in the corresponding moves in superstate sampling.

For the kinds of moves which we want to use to change the outer state, we randomly choose the outer state to be proposed from a suitable set (cf.~Fig.~\ref{g_config}). This involves first following essentially the same procedure for a move as in superstate sampling, which allows us to obtain an outer superstate proposal which we could call the target superstate. From this target superstate, we randomly choose a state to propose as outer state, in our specific implementation according to the distribution in Eq.~(\ref{eq:outerstateproposal}). In this optimized probability distribution for the outer state proposal, we take only the eigenenergies of the different possible outer states into account, as we have found this to be a useful way to increase the sampling efficiency.

Let us take a look at some detailed examples for these latter kinds of moves: When the outer pair insertion (cf.\ Sec.~\ref{sec:sstoldmoves}) depicted by Fig.~\ref{fig:innerouter} is performed in state sampling, the initial configuration depicted in the middle is of course extended by the specification of one outer state $s \in \mc{A}$, and one of the states contained in $\mc{E}$ is randomly selected as the outer state of the proposed new configuration depicted below it. Similarly, in the $\tau$-shift move (cf.\ Sec.~\ref{sec:ssttaushift}) depicted by Fig.~\ref{fig:taushift}, the initial configuration depicted on top is extended by the specification of an outer state $s \in \mc{A}$ and one of the states contained in $\mc{C}$ is randomly chosen as the outer state of the proposed configuration. In any further global moves with no particular connection to the superstate sampling technique, we chose to randomly propose one of the contributing superstates. Therefore when they are performed in state sampling, we randomly choose one of the states contained by the contributing superstates---the only case in our implementation where outer states from more than one different superstate could be proposed in one otherwise identical move.

Proposing one of the possible states with uniform probability, however, causes a lower acceptance rate compared to the analogous moves in superstate sampling, where, e.g., the $\tau$-shift move even has acceptance rate 1.
Therefore, we employ a more efficient strategy:
Since close to $\tau \! =\!0$ there is always just the outer state $s$ propagating with its eigenenergy,
we can make the procedure more efficient by ``transferring'' the time evolution of this state from
the acceptance to the proposal probability; i.e., we include it in the proposal probability so that in the standard Metropolis acceptance probability formula, it cancels with the equivalent factor in the weight of the configuration.  This proposal probability weighting allows us to
incorporate our prior (or easier to calculate) knowledge to
avoid wasting time on proposals that would likely be rejected: For a well chosen proposal probability, i.e., one close to the actual weight distribution, we raise the acceptance probability for all outer states since those outer states that would be rejected more often with uniform proposal probability are simply proposed less often. To include the aforementioned time evolution close to $\tau = 0$, we use the proposal probability
\begin{equation}
  \label{eq:outerstateproposal}
  p_\text{prop}(s) = \frac{\exp(-(E_s - E_0) \cdot (\tau_f + \beta-\tau_l))}{\sum_{k \in \mathcal{T}} \exp(-(E_k - E_0) \cdot (\tau_f + \beta-\tau_l)) },
\end{equation}
where $\mathcal{T}$ contains all states from all outer superstates
that may be proposed and $\tau_f$ and $\tau_l$ are, respectively, the imaginary times of the first and last operator after the move. In this
way, the acceptance rate for outer state changes is significantly
increased and, e.g., reaches about $50\%$ for the global $\tau$-shift move
in typical calculations. The best choice of proposal probability for such an optimization depends on the weight we expect: In this case, we essentially assume that the potentially excited state at $\tau = 0$ will be brought closer to the ground state by the impurity operators, since if we expected the energy to stay at the level of the outer state, we could choose a better proposal probability assuming propagation over the entire imaginary time with the energy of the outer state (which corresponds to replacing $\tau_f + \beta-\tau_l$ by $\beta$ in the proposal probability).

\section{Speed-up quantification}
\label{sbenchmarks}

To check the correctness of the results, we use a two-orbital model
with a Kanamori interaction, i.e., density-density, pair-hopping, and
spin-flip terms, and a finite number of bath sites for which we have a
reference solution obtained using exact diagonalization. Both the
reference self-energy as well as one calculated using our CT-HYB
solver are shown in Fig.~\ref{g_siw} (Appendix~\ref{EDbenchmark}).

In order to analyze the performance, we use a five-orbital Hamiltonian
modeling a realistic transition-metal impurity on the surface of a
metal\cite{CNoteModel}
with both the full (spherically symmetric) Coulomb tensor as well as one
derived from the same interaction matrix restricted to Kanamori-like
terms only. We perform all calculations using
\textit{w2dynamics}\cite{w2dynamics}, with either the implementation
of conventional sampling\footnote{Specifically, a matrix-vector
  implementation of CT-HYB with time evolution in eigenbasis. Quantum
  number checking (cf.\ Sec.~\ref{s_cthyb}) is used to avoid the
  calculation of zero contributions to the local weight using a
  partitioning of the Hilbert space into superstates using conserved quantities for Kanamori-like
  interaction\cite{parragh-conserved-2012} and additionally automatic
  partitioning\cite{haule-quantum-2007,triqs} for full Coulomb interaction. We do not
  use outer truncation of the local trace, sliding-window-style
  local updates, tree algorithms, or other optimizations even if mentioned
  in the introduction unless explicitly stated.}
  found in older versions, or 
superstate and state sampling proposed here; other than the
sampling method, there are no differences with significant performance
impact between the calculations. To quantify the performance improvement we compare
the sampling rates, i.e., the raw amount of generated (correlated)
samples per time. For the autocorrelation time we found a minor increase of about 10 $\%$
for superstate sampling, but only about 3 $\%$ for state sampling
as compared to the old sampling method.
This can be considered negligible in
comparison to the speed-up factors. The mean sign is about $1.0$
for the model with Kanamori interaction using both conventional and
superstate sampling and about $0.98$ using state sampling. 
For the model with full Coulomb interaction the sign is significantly less than $1$
in all cases, and it is slightly smaller with state
sampling than with the other methods\footnote{We used about the same
amount of CPU time for all sampling methods for this example, so due
to the worse performance of the conventional implementation its error
is considerably larger. While the error in this example is thus
sufficiently large to allow other conclusions about the relative sign
of the methods in some temperature ranges, data from many other
calculations we did with all methods not specifically for the purpose
of this article strongly indicate equal signs for conventional and
superstate sampling and a sign closer to zero (with model-dependent
extent) for state sampling.}
For the calculations used to measure the speed up factors, the absolute values of the mean sign
for all three methods can be found in Fig.~\ref{fig:sign}.

\begin{figure}[htbp]
  \raggedright
  \hspace*{0.8em}\includegraphics[scale=0.97]{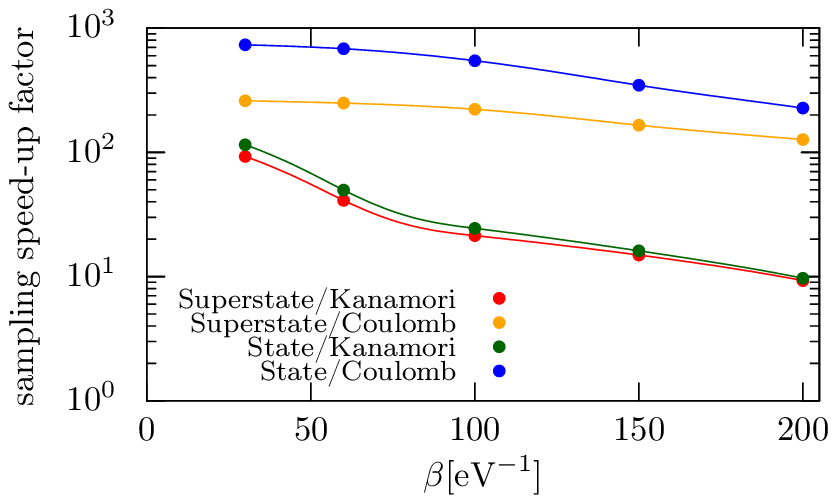}\\%
  \includegraphics{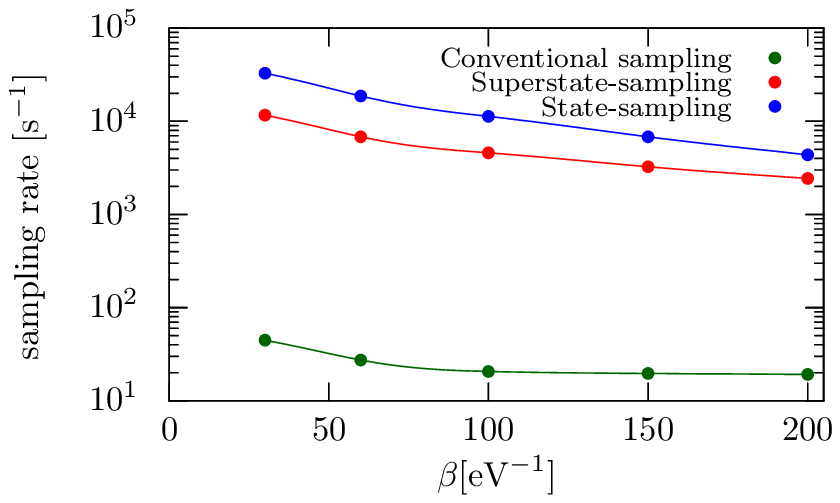}
  \caption{Comparison of Monte Carlo sampling speed, i.e.\ number of individual, mostly local, updates per CPU time, not including measurements, for a five-orbital system with Kanamori or
    Coulomb interaction and cubic interpolation curves plotted
    logarithmically. Top panel: speed-up factors of the sampling
    methods compared to conventional sampling for Kanamori and
    Coulomb interaction. Bottom panel: speed of all the sampling methods
    for Coulomb interaction.}
  \label{fig:speedup}
\end{figure}

\begin{figure}[htbp]
  \centering
  \includegraphics{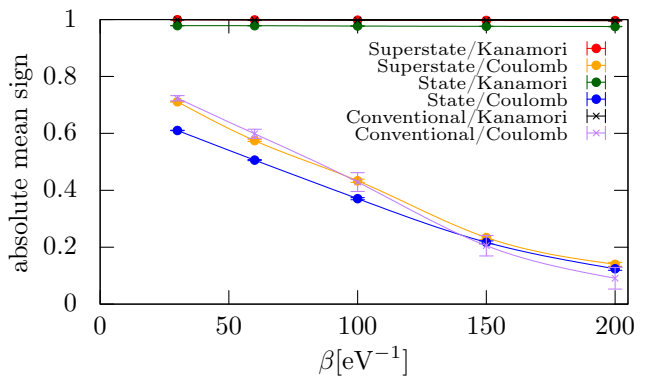}
  \caption{Absolute value of the mean sign using our sampling
    methods for a five-orbital system with Kanamori or Coulomb
    interaction with cubic interpolation. The sign obtained in
    simulations using conventional sampling is usually not better than
    with superstate sampling, which we confirmed for the case with
    Kanamori interaction. Note that the error of the sign for the model with
    Coulomb interaction is considerably larger using conventional
    sampling than using other algorithms. The graph also shows that the sign obtained in
    state sampling is only slightly worse that that obtained using
    the other methods.}
  \label{fig:sign}
\end{figure}

Figure \ref{fig:speedup} shows the achieved speed up factors (top panel)
and the absolute sampling rates measured in simulations (bottom panel)
of the impurity model with Coulomb interaction.
We obtain a speed up of the Monte Carlo sampling up to a factor of about 700 
in the considered temperature range, depending on the used sampling method, temperature,
and interaction. 
Remarkably, the speed up of our five-orbital example was never smaller than 100 for the most arduous case, i.e., the full interaction.
The reason why the speed up factors for the Coulomb
interaction are larger under otherwise equal conditions is that a larger
number of superstates contribute on average in this case. A general observation
is that the speed up decreases with decreasing temperature because the
number of superstates contributing to the trace decreases with
decreasing temperature as more operators tend to cause more quantum
number violations. Therefore only the quantum number checking
can be avoided at lower temperatures, whereas at higher temperatures
other trace contributions are present for which matrix-vector products
need not be calculated any more.

Yet, there is still a noticeable speed up
even at lower temperatures where quantum-number checking takes a large amount of the total time \footnote{In this work we consider the computational cost of quantum-number checking in \emph{w2dynamics} \cite{w2dynamics} with decreasing temperature. The magnitude of this speed up might of course be lower in other
implementations.}. As the speed-up affects only the trace calculation,
it will also continue to decrease with $\beta$ because the computational complexity
of the bath determinant (which scales with $\beta^3$) is asymptotically larger than that of the
trace calculation.

Another noticeable feature is the bigger advantage of state sampling
over superstate sampling for Coulomb interaction, which is due to the
larger average size of the outer superstates in that case. The amount
of calculated matrix-vector products is reduced by approximately that
factor in state sampling compared to superstate sampling, as only one
of the outer states is chosen in the former case. This optimization is
only advantageous for a matrix-vector solver like ours, as additional
outer states can be included at negligible further cost if the entire
product of the operator matrices for a specific outer superstate has
been calculated, cf. Sec.~\ref{sec:statesampling} and Appendix~\ref{sec:mmmv}. A similar optimization is also possible without
splitting configurations in the superstate sampling method by using
the cyclical invariance of the trace and starting the trace
calculation at a $\tau$ where the superstate of the configuration is
smaller than the outer superstate of the configuration, but this can
interfere with time savings from caching of intermediate state vectors
and even the size of the smallest superstate may be greater than 1.
More details on the methods can be found in Ref.~\onlinecite{kowalski-improved-2017}.

In conclusion, we find that superstate sampling improves performance without significant drawbacks to such an extent that it should always be preferable to the conventional sampling method. Especially in simulations with few good quantum numbers, state sampling can provide an additional speed-up, though it can also impact the mean sign. In our examples, the speed-up in the case with full Coulomb interaction is big enough to clearly outweigh the marginally reduced sign, but this may depend on characteristics of the system and the implementation.

\section{Application: the spin-freezing crossover}
\label{asdfspin}

In Ref.~\onlinecite{werner-spin-2008} Werner, {\it et al.} applied DMFT to a model with three degenerate orbitals and rotationally-invariant Coulomb interaction. Upon changing the filling $n$, they identified a sharp change in the qualitative behavior of the local spin susceptibility. 
For small fillings, the latter becomes rapidly small at large imaginary times $\tau$, as in a standard metal. Approaching the Mott transition at $n\!=\!3$, it starts instead to closely resemble that of an atomic insulator, i.e., it seems to become essentially constant in $\tau$. This is surprising, as it happens for fillings still on the metallic side, before reaching the metal-insulator Mott-Hubbard transition. 
The sudden loss of coherence was interpreted as an abrupt crossover -- or even a true quantum phase transition (``spin-freezing'') -- to a bad metal characterized by violations of the Fermi-liquid properties, along a line in the zero-temperature $n$-$U$ phase diagram.
\begin{figure}[th]
   \centering
   \includegraphics{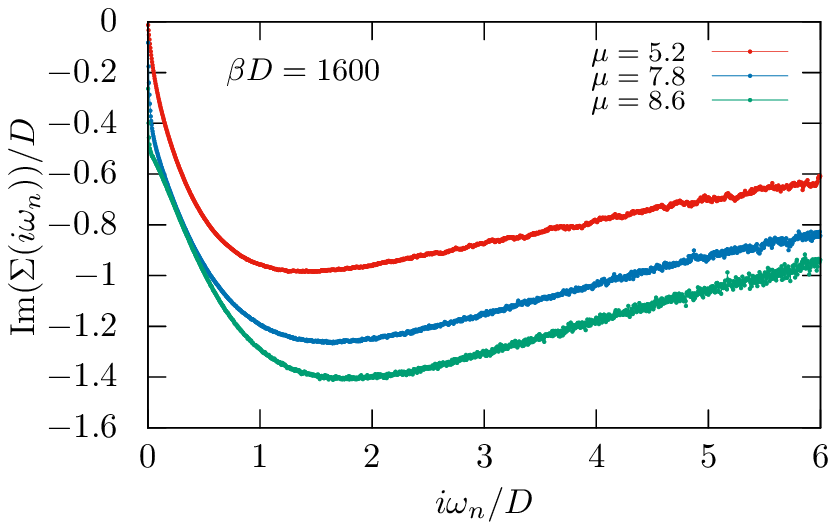}\\
   \includegraphics{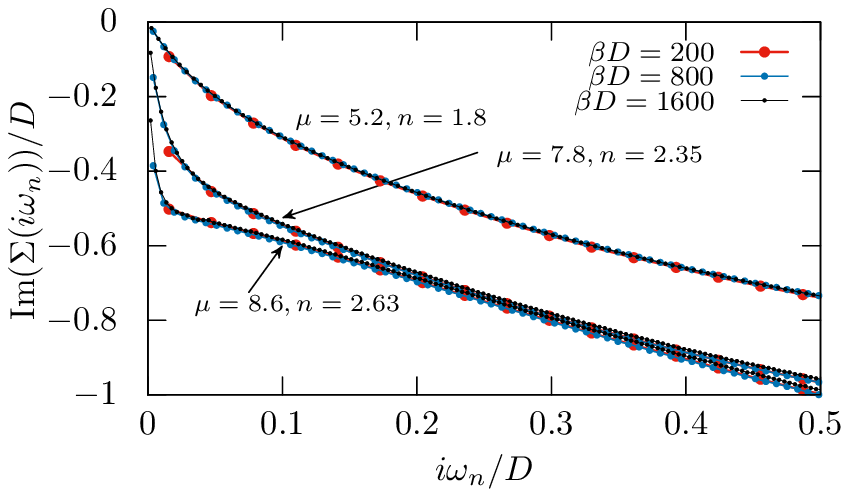}
\caption{Imaginary part of the self-energy on the Matsubara axis with $U/D\!=\!4$, $J/D\!=\!2/3$. Upper panel: three different chemical potentials at low temperature $\beta D=1600$. The corresponding fillings are reported in the lower panel and represent values across the spin-freezing crossover (see inset to Fig.~\ref{g_chiloc}). A close up of the low-frequency region is shown in the lower panel together with the evolution with temperature. The Fermi liquid behavior for $n\!=\!2.63$ ($\mu\!=\!8.6$), i.e., deep in the ``spin-freezing region'', is recovered only when considering the lowest temperatures. It indeed occurs at such a low-energy scale that its existence is not foreseeable by extrapolating the data at, e.g., $\beta D=200$.} 
   \label{g_siws}
\end{figure}

An independent analysis showed, however, convincing evidence that the same model remains in a metallic phase away from integer fillings~\cite{de-medici-janus-faced-2011}. What changes upon getting close to the Mott transition is the coherence of the quasiparticle excitations. The ``spin-freezing'' is therefore a finite-temperature---though surprisingly rapid---crossover to a ``bad-metal'' rather than a $T\!=\!0$ phase transition.
As long as the DMFT self-consistence does not lead to a gap in the spectral function of the bath of the corresponding Anderson impurity model, the solution at zero temperature must indeed be a Fermi liquid. This conclusion was already demonstrated more rigorously in Refs.~\onlinecite{de-leo-non-fermi-2004,leo-spectral-2004}.

Yin, Haule, and Kotliar~\cite{haulePRB86} and, immediately after them, Georges, de' Medici, and Mravlje~\cite{georges-strong-2013} performed CT-HYB calculations of hitherto unprecedented efficiency, reaching temperatures 1000-1500 times smaller than the half bandwidth (see Fig.~6 and Fig.~7 in the two papers, respectively).
The focus was on the functional dependence of $\text{Im} \Sigma(i\omega_n)$ which, even after the spin-freezing crossover, was shown to follow a Fermi-liquid scaling at extremely low frequencies, visible at these temperatures.
A recent study nailed this down using an advanced multiorbital numerical renormalization group solver \cite{stadler-hundness-2018}.

Similar types of crossovers have been discussed in the presence of spin-orbit interaction~\cite{kim-j-2017} always showing the change of behavior in the local spin susceptibility at a fixed temperature. Yet, one would like to unambiguously demonstrate that this is actually the physics of a crossover from a good to a bad metal with a coherence temperature that becomes fairly small upon approaching the Mott transition at half filling. 

Here, we consider the same model of Ref.~\onlinecite{werner-spin-2008} as a function of doping focusing in particular on the temperature dependence. Using the CT-HYB implementation of \emph{w2dynamics}\cite{w2dynamics}, featuring both superstate sampling and sliding window sampling\cite{shinaoka-hybridization-2014}, we are able to obtain a clear picture of the temperature evolution of the local spin response, identifying a coherence scale, even deep into the ``spin-freezing region''.
The quantities of interest are the electronic self-energy $\Sigma(i\omega_n)$ and the static local spin susceptibility 
\begin{equation}
\label{chi_loc}
\chi_{\mathrm{loc}}^{\omega=0}(T) =\int_{0}^\beta \! \mathrm{d}\tau \, \chi_{\mathrm{loc}}(\tau),
\end{equation}
i.e., the $\omega\!=\!0$-Fourier component of the spin-spin response function $\chi_{\mathrm{loc}}(\tau)=g^2 \sum_{ij} \left< S_z^i(\tau)S_z^j(0) \right>$ (with $i$ and $j$ running over the three orbitals). The half bandwidth of the semicircular noninteracting density of the states of each orbital is $D$ (corresponding to 2$t$ in Ref.~\onlinecite{werner-spin-2008}). The coupling constants of the three-orbital Kanamori interaction are the usual Hund-$J$ and Hubbard-$U$ (with $V\!=\!U-2J$) at fixed $J/U\!=\!1/6$ ratio. 

We first focus on the imaginary part of the Matsubara self-energy $\text{Im} \Sigma(i\omega_n)$ for three different fillings at the lowest temperature $T\!=\!1/\beta\!=\!D/1600$. As shown in Fig.~\ref{g_siws}, for filling $n\!=\!1.8$ the extrapolation of $\text{Im} \Sigma(i\omega_n)$ for $i\omega_n \!\! \rightarrow \! 0$ is not dramatically influenced by the temperature. This indicates that the system has reached a coherent Fermi-liquid state and further lowering $T$ does not change the shape of $\text{Im} \Sigma(i\omega_n)$ but only makes the Matsubara frequencies denser, remaining on the same ``straight'' line. This is a manifestation of the so-called ``first-Matsubara'' rule~\cite{chubukov-first-matsubara-frequency-2012,pourovskii-electronelectron-2017,hausoel-local-2017}, according to which a $T^2$ scattering rate characteristic of a Fermi liquid gives rise to a linear-in-$T$ value for $\text{Im} \Sigma(i\omega_{n=0})$.

The situation is drastically different for larger values of $n$. Note that at this $U$, $n\!=\!2.35$ and $2.63$ had been already assigned to the ``spin-freezing region'' in the original paper by Werner, {\it et al.} (see inset to Fig.~\ref{g_chiloc}). At these fillings the low-frequency part of $\text{Im} \Sigma(i\omega_n)$ is highly nonlinear and it is clear that to recover linearity one has to consider the lowest temperatures (and probably even lower than $T\!=\!D/1600$ at $n\!=\!2.63$). This unambiguously hints at a sudden drop of the Fermi-liquid coherence temperature upon increasing the filling $n$.

\begin{figure}[h]
   \centering
\includegraphics{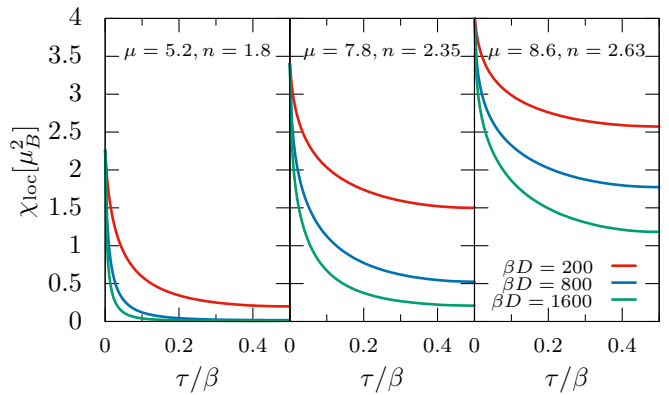}
\caption{Local spin susceptibility for three different fillings across the spin-freezing crossover. While the instantaneous moment at $\tau \! = \! 0$ gets larger upon increasing the filling, the value at $\tau \! = \! \beta/2$ can be made arbitrarily small upon reaching low enough temperatures. This supports the picture of a Fermi liquid with progressively lower and lower coherence scale.}
   \label{g_chis}
\end{figure}
An inspection of the local spin susceptibility confirms that the physics at $n\!=\!2.35$ and $2.63$ is not qualitatively different from the good-metal fillings but it is just the result of a strong renormalization of the coherence properties.
The results are shown in Fig.~\ref{g_chis}. For an atom, $\chi(\tau)$ is perfectly flat independently of the temperature, so that its integral from 0 to $\beta$ is proportional to $\beta$ (Curie law). For a Fermi liquid, its shape instead has to change with temperature in such a way that its integral gives a constant Pauli susceptibility. Even though the speed of the decay for $n\!=\!2.35$ and $2.63$ is greatly reduced compared to $n\!=\!1.8$ (in agreement with Werner {\it et al.}), a pronounced temperature dependence of $\chi(\tau)$ is present also for the larger fillings, revealing the Fermi-liquid properties. 

\begin{figure}[h]
   \centering
   \includegraphics{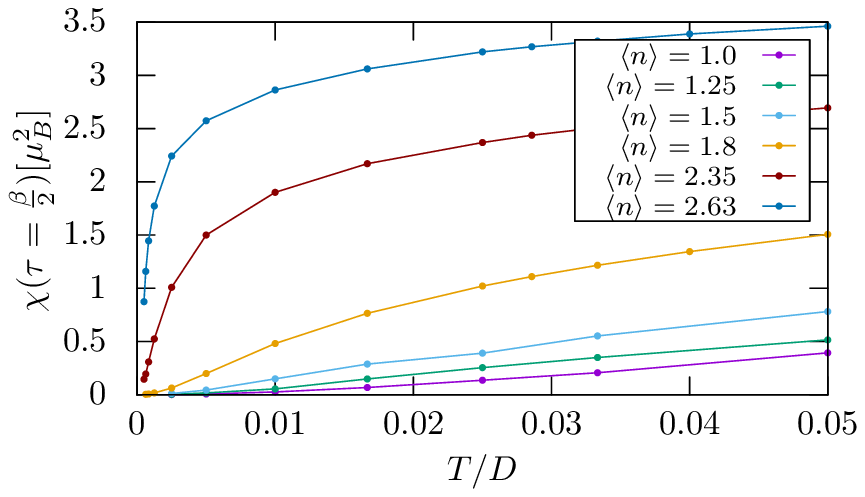}\\
   \includegraphics{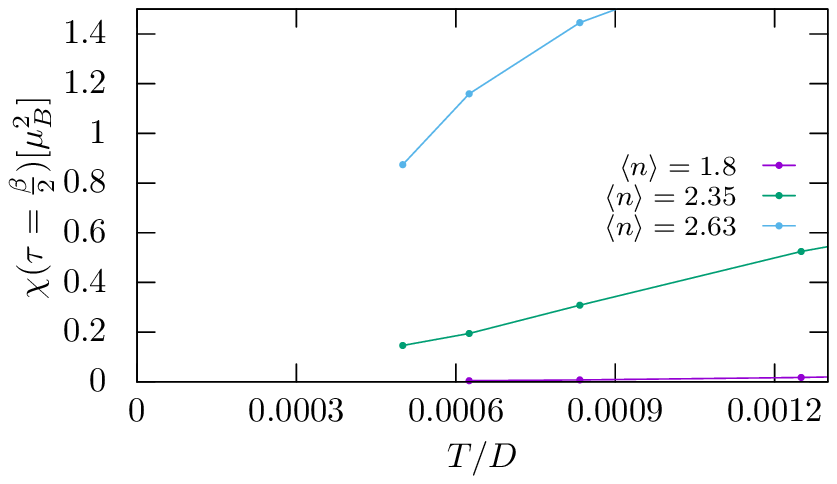}
   \caption{Behavior of $\chi(\tau \! = \! \beta/2)$ for different temperatures and fillings. The lower panel contains a close-up of the three fillings shown in Fig.~\ref{g_chis} across the spin-freezing crossover.}
   \label{g_betahalf}
\end{figure}
To conveniently represent the evolution with the filling we look at the value of the susceptibility at $\tau \! = \! \beta/2$: In the Fermi-liquid case this has to go (quadratically) to 0 upon reducing $T$. 
The coherence temperature can be estimated for instance from the inflection point of $\chi(\tau\! = \!\beta/2)$ (see also Fig.~\ref{g_tcoh}). 
By looking at Fig.~\ref{g_betahalf} one immediately understands how the filling efficiently reduces the temperature scale at which $\chi(\tau \!= \!\beta/2)$ approaches 0 and hence how it makes the coherence temperature drop. At the same time our results reveal how the physics of this model is qualitatively the same at all metallic fillings. The difference between the curves for different $n$ is indeed only the velocity of the renormalization and the temperature scale at which the Fermi-liquid scaling is recovered. 

\begin{figure}[h]
   \centering
\includegraphics[width=\linewidth]{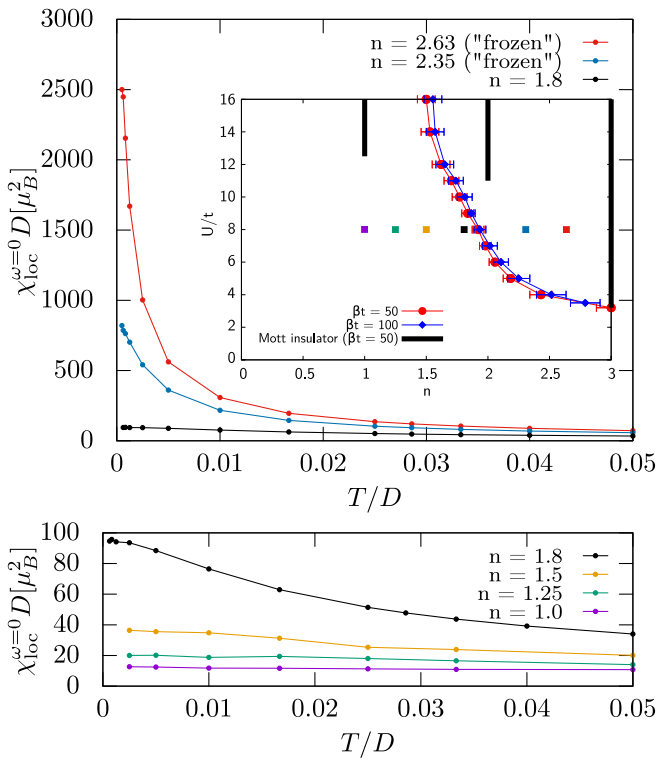}
\caption{Temperature dependence of $\chi_{\mathrm{loc}}^{\omega=0}(T)$ for different fillings (indicated by square markers of the same colors in the inset taken from Ref.~\onlinecite{werner-spin-2008}). The low-$T$ Pauli-like behavior is visible in the ``nonfrozen'' regime (lower panel). At higher temperature there is a crossover to Curie-Weiss physics. The latter gets more and more dominant in the ``frozen'' regime (upper panel; upper right region of parameter space in the inset).}
   \label{g_chiloc}
\end{figure}
The crossover from Curie to Pauli upon reducing the temperature can also be visualized in Fig.~\ref{g_chiloc}. Upon approaching the spin-freezing crossover the Pauli behavior gets progressively pushed to lower and lower temperatures. In the curves at $n\!=\!2.63$ for instance it is clear that even lower temperatures would be needed to fully resolve it. 
For this reason it is hard to unambiguously prove that deep into the ``spin-freezing region'' the coherence scale is actually exponentially small. 
Even in the good metal region a precise estimate of the crossover temperature $T_\text{coh}$ is not a trivial task. First of all there is a dependence on the observable one is focusing on. Second, even by looking at the same physical quantity, different criteria give somewhat different answers. 

\begin{figure}[th]
   \centering
\includegraphics{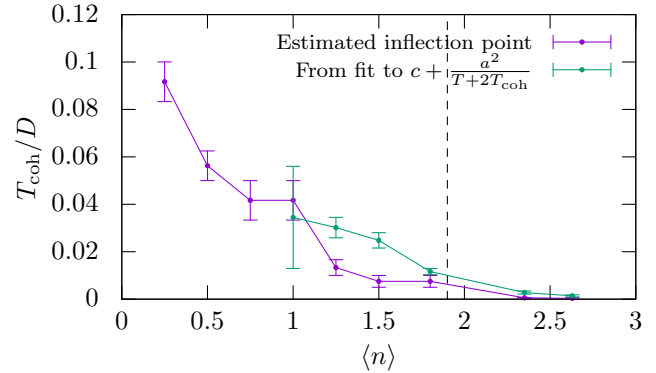}
\caption{Coherence temperature $T_{\mathrm{coh}}$ determined with two different methods: with the inflection point of the curves shown in Fig.~\ref{g_betahalf} and by fitting the high temperature region with Wilson's formula \cite{wilsonRMP,hausoel-local-2017}. We have not been able to obtain good results using the latter method at the lowest densities.}
   \label{g_tcoh}
\end{figure}
In Fig.~\ref{g_tcoh} we quantify $T_\text{coh}$ via two approaches based on $\chi(\tau \!= \!\beta/2)$. 
These two ways of quantitatively estimating $T_\text{coh}$ give compatible results though affected by sizable errors.
Furthermore, even though the crossover from the high-temperature $\sim 1/T$ Curie to the Pauli region becomes relatively sharp (i.e., in principle more easily identifiable) after crossing the ``spin-freezing'' crossover line, we have difficulties in precisely determining $T_\text{coh}$, as the latter is very small there and we do not have many susceptibility data points at such low temperatures. 
Nevertheless, the existence of a \textit{sudden} drop of $T_\text{coh}$ approaching half-filling, as in fact pioneered by Werner {\it et al.} in Ref.~\onlinecite{werner-spin-2008}, is undoubted. 
Similar conclusions are corroborated by high-precision numerical renormalization group studies, published in Refs.~\onlinecite{stadlerPRL,horvatPRB}. 
The reason why this crossover is so sharp, as well as its shape in the doping-$U$ diagram, are not fully clear yet\cite{lucaLecture}.

In real materials, the position of the coherence scale can be strongly influenced by several factors, such as the nonlocal hybridization between orbitals, absent in the model Hamiltonian studied here. One of these factors has been also identified in the presence of sharp peaks in the noninteracting density of the states \cite{hausoel-local-2017}, something often coexisting with the many-body physics in strongly correlated materials.

\section{Conclusions}

We have shown that the sum over all impurity eigenstates of the local problem in CT-HYB can be 
divided into smaller pieces, and sampled individually. This fragmentation leads to a remarkable gain in the algorithm's efficiency, to some extent against the general intuition. This is due to the exponential character of the imaginary time evolution $\mathrm{e}^{-H_{\mathrm{loc}}\tau}$, 
which very sensitively damps the amplitudes of high energy excitations. 
Acting on the core of the exponential scaling in CT-HYB we manage to achieve speed-up factors of the order of $10^3$, with essentially no worsening of the average sign. 
Additional work has to be carried out in order to show whether the impact of the exponential scaling of the local problem can be reduced further by employing methods based on our ideas.

The speed up figures have been obtained for a five-orbital model with full-Coulomb interaction, representing physically relevant situations such as realistic transition-metal impurities deposited on metallic substrates. 
We also discussed the well-known spin-freezing crossover obtained in three-orbital Hubbard-model calculations. Reaching very low temperature allows us to quantify the coherence temperature and the recovery of the Fermi-liquid properties of the self-energy, even when this physics is pushed to very low scales by the proximity to the Mott transition.

\section*{Acknowledgments}
We thank M.~Aichhorn, F.~Assaad, M.~Capone, L.~de' Medici, K.~Held, M.~Karolak, G.~Kotliar, O.~Parcollet, H.~Strand, A.~Toschi and N.~Wentzell for useful discussions.
This work was supported by the DFG through SFB1170 ``Tocotronics''. G.~S.\ and A.~H.\ thank the Flatiron Institute, a division of the Simons Foundation, for the fruitful scientific exchange and the kind hospitality. We gratefully acknowledge the Gauss Centre for Supercomputing e.V. (www.gauss-centre.eu) for funding this project by providing computing time on the GCS Supercomputer SuperMUC at Leibniz Supercomputing Centre (www.lrz.de).

\appendix

\section{Proof of ergodicity of $\tau$ shift moves}
\label{s_proof_of_ergodicity_tau_shifts}

To prove ergodicity, rigorously at least for the case of density-density interaction, by connecting two configurations differing only in their outer superstates, we first take an arbitrary configuration $\mathcal C$ and connect it to the
``empty'' configuration, which has no operators: For an arbitrary outer superstate, consider the occupation number basis states contained by the superstates. To get from one superstate to another, impurity operators need to be applied such that a state from the first superstate would be transformed into a state from the second one. To do this in a simulation, perform inner pair moves to build up this sequence of operators right after $\tau$; the second operator from each pair can be placed anywhere resulting in a configuration of nonzero weight, with one simple possibility being $\beta - \tau$ if the first is placed at $\tau$. Now, perform one tau-shift move with $\Delta \tau$ being the point right after the last operator of the inserted sequence, and the outer superstate is changed from the old, ``first'' one to an arbitrary ``second'' one. By removing all operator pairs in the opposite order, an ``empty'' configuration with an arbitrary outer superstate is reached.  We can now run the procedure in reverse to build up a new configuration $\mathcal C^\prime$.  This implies ergodicity,  as any two configurations $\mathcal C$ and $\mathcal C^\prime$ are connected by a finite number of moves.

It is simple to demonstrate that ergodicity is not lost if outer pair moves are replaced by global $\tau$-shift moves or vice versa: An outer move can be replaced by a sequence of a $\tau$-shift move, an inner move, and the $\tau$-shift move inverse to the first one. In the first move, $\Delta \tau$ just needs to be sufficiently big to change the order of the operator pair that would be affected by the outer move in $\tau$, then it can be performed as an inner move instead because the section of the trace that is changed has entirely been moved inside. A $\tau$-shift move can be replaced by a sequence of outer and inner moves: Remove all operators in pairs using outer and inner moves such that an operator-free configuration with the desired outer superstate is obtained, then put all operator pairs shifted by $\Delta \tau$ back in using outer and inner moves.

As a final remark, it should be stressed that even though a $\tau$-shift move has acceptance 1 in superstate sampling, the acceptance of the outer pair moves remains smaller than that of the inner pair moves. This is due to the fact that outer moves are connected to inner moves by \textit{specific} $\tau$-shifts that have a proposal probability smaller than 1.

\section{Invariance of the bath weight under global $\tau$-shift move}
\label{s_bath_proof}

For the proof of the invariance of the bath determinant (adapted
from Ref.~\onlinecite{kowalski-improved-2017}), let us
consider the form of the hybridization matrix elements with time
ordering along both dimensions,
\begin{equation}
  \label{eq:hybmatelm}
  \Delta_{ij} = \Delta \left( \tau_i - \tau^\dag_j \right),
\end{equation}
where $\Delta\left(\tau\right)$ is the hybridization function, an
antiperiodic function with period $\beta$, $\tau_i$ the imaginary time
of the $i$-th annihilator (ordered by imaginary time), and
$\tau^\dag_j$ the imaginary time of the $j$-th creator. The number of
annihilators and creators shifted across $\tau = 0$ by the move will
be denoted as $N_A$ and $N_C$ in the following.

Due to the $\tau$-shift move, the imaginary time of
operators with $\tau < \Delta \tau$ will be transformed as
$\tau \rightarrow \tau + \beta - \Delta \tau$ and that of other
operators as $\tau \rightarrow \tau - \Delta \tau$. The arguments of
the hybridization functions are only time differences in which the shift
parameter $\Delta \tau$ always cancels, but in cases where exactly one
of the two operators had a $\tau < \Delta \tau$, the corresponding
matrix element changes its sign.

Additionally, since the ordering of
the operators is cyclically permuted, the rows and columns are
cyclically permuted such that the $N_A$ first rows become the $N_A$
last rows and the $N_C$ first columns become the $N_C$ last rows,
where $N_A$ is the number of annihilators with $\tau < \Delta \tau$
and $N_C$ the number of creators with $\tau < \Delta \tau$. A cyclic
permutation that moves every column or row exactly one position toward
the front (``wrapping around'' from the beginning to the end) is
equivalent to swapping adjacent columns or rows $k-1$ times, where $k$
is the size of the matrix in that dimension (as the hybridization
matrix is a $k \times k$ matrix for hybridization expansion order
$k$).

Each swap causes the determinant to change its sign, and the
sign change of matrix elements where only one operator wrapped around
the end is equivalent to multiplying all wrapped rows and columns by
$-1$, where every multiplication of a column or row causes the
determinant to change its sign as well. In total, expressed using the
number of wrapped operators $N_\text{wrap} = N_A + N_C$, the
determinant thus accumulates an additional factor of
${\left( {\left( -1 \right)}^{k-1} \right)}^{N_\text{wrap}} \cdot
{\left( -1 \right)}^{N_\text{wrap}} = {\left( -1 \right)}^{k
  N_\text{wrap}}$, where the matrix size $k$ is the expansion order.

This extra factor is compensated by the sign that is incurred due to
time ordering. That the change of this extra sign is equal to the
factor acquired by the determinant may be proven by considering the
amounts of permutations necessary to restore the ordering after
performing a $\tau$-shift move that wraps exactly one operator around the
origin. From this, the general case follows.

\section{Invariance of the local weight under global $\tau$-shift move}
\label{s_trace_proof}

To prove the invariance of the local weight in superstate sampling under a $\tau$-shift move, we use the cyclic
invariance of the trace (adapted from
Ref.~\onlinecite{kowalski-improved-2017}). Due to the way superstates are chosen by
definition we know that if the trace is restricted in such
a way that nonzero components are left for only one superstate at any $\tau$, the
result will be the same as if done so everywhere. Our local weight is effectively the conventional local weight with
such a restriction applied at $\tau = 0$ and $\tau = \beta$, and if 
projection operators $P^{(x)}$ are inserted onto the outer superstate $x$ at
the beginning and end of the product of time-evolution, creation and
annihilation operators corresponding to current configuration, it can be
written as a proper trace:
\begin{align}
  \label{eq:sstweightproj}
  w_\text{loc} &= \sum_{s \in x} \bra{s}\hat{\mathcal{C}}\ket{s}\\
               &= \tr\nk{P^{(x)} \hat{\mathcal{C}} P^{(x)}}.
\end{align}

As the time evolution does not mix states from different superstates,
the superstate projection operator commutes with all time-evolution
operators,
\begin{align}
  \label{eq:sstexphcomm}
  \ek{\exp(-\tau \hat{H}), P^{(x)}} = 0,
\end{align}
for any superstate $x$ and any $\tau$. Because the creators and
annihilators map each source superstate to one unique target
superstate and vice versa, a projector onto a superstate $x$
on one side of an annihilator or creator can be replaced by a
projector onto the superstate $y$ which the operator maps $x$ to on
the other side of the operator:
\begin{align}
  \label{eq:sstcacommute}
  d^{(y|x)} P^{(x)} = P^{(y)} d^{(y|x)},
\end{align}
where the mapping of superstates relevant for the specific case is
given in parentheses in superscript with the meaning that applying
$d^{(y|x)}$ to a state vector (to the right) in the subspace of
superstate $x$ will produce a state vector in the subspace of
superstate $y$.

Using these two relations, we can commute the projectors all the way
through the product to any other point and also to any imaginary time
by splitting time evolution at that time into two consecutive
time evolutions if necessary. The projectors will not
necessarily be projectors onto the old outer superstate any more, but onto
the superstate that can be found at that point in the sequence for the
current configuration. After commuting both projectors to the position
in the product corresponding to the imaginary time $\Delta \tau$, the
product can be cyclically permuted such that one of the projectors
ends up at each end of the trace. It is then equivalent to the local
weight of the superstate sampling configuration after a $\tau$-shift by
$\Delta \tau$. This shows that the local weight does not change after
a $\tau$-shift move.

\section{Compatibility of state sampling with implementations}
\label{sec:mmmv}
Whether the performance of state sampling actually exceeds that of superstate sampling depends on the type of CT-HYB implementation it is used with. When reviewing standard CT-HYB in Sec.~\ref{s_cthyb}, we simply expressed the weight of a configuration as the trace of the product of the corresponding impurity operators. Typically, there are two ways to calculate it: In a matrix-matrix implementation, the matrices representing the impurity operators are multiplied with one another and the trace is obtained from the diagonal elements of the total matrix product. In the matrix-vector flavor, we instead explicitly perform a sum over the outer states: For each outer ket-vector, we repeatedly calculate the matrix-vector product with the operators, starting from the first all the way through to the outer bra-vector, with which we finally compute the scalar product.

By decomposing the problem into superstates (i.e., block diagonalizing the Hamiltonian and choosing the blocks such that the operators connect them in a one-to-one way), we simplify the calculation of the trace in that the outer states (per superstate) follow independent, noncrossing paths through the superstates. As a result, in the matrix-vector implementation one can just reduce the size of the initial operator matrices. In a matrix-matrix implementation the trace of the product can be decomposed into the sum of traces per outer superstate, which introduces the explicit summation whose terms can be calculated using matrices of reduced size as well.

If we consider such an implementation as the starting point, we just have to restrict the summation to one outer superstate in either case to implement superstate sampling. While we can obviously reduce the number of needed calculations by restricting the outer sum to one state only in the matrix-vector implementation, i.e., implementing state sampling, there is no way to beneficially implement this in a matrix-matrix algorithm. We could use only one of the diagonal elements of the resulting products, but this does not make the product calculation simpler and therefore does not improve performance but would only waste the other contributions that could be included at negligible further cost.

\section{Exact diagonalization cross-check} \label{EDbenchmark}

Here we show that the results of the CT-HYB results with superstate sampling agree with an exact-diagonalization benchmark.
\begin{figure}[hbtp]
   \centering
   \includegraphics[width=\linewidth]{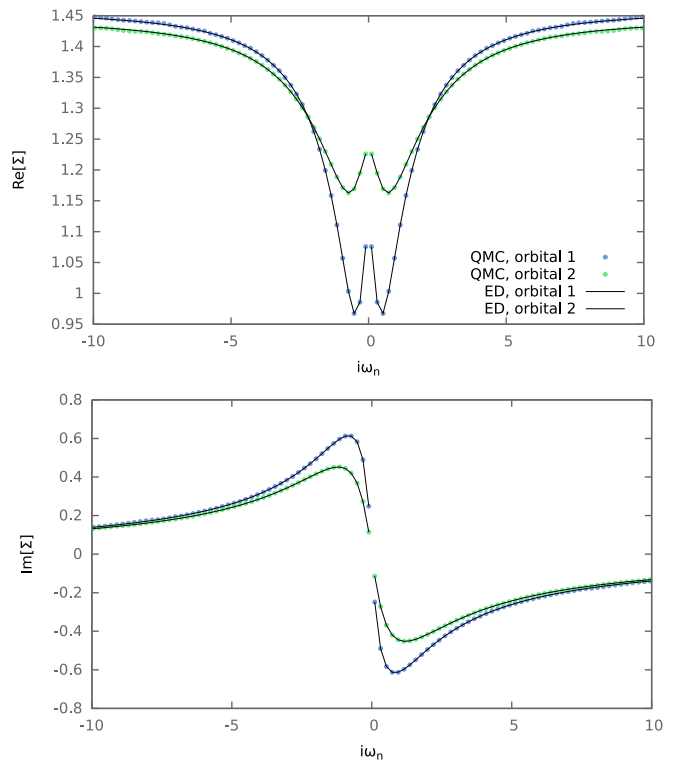}
   \caption{Comparison of the CT-HYB self-energy calculated using the
     superstate sampling method (points) to exact diagonalization (lines) for a
     two-orbital system with Kanamori interaction.\label{g_siw}}
\end{figure}

\bibliography{main}

\begin{thebibliography}{39}%
\makeatletter
\providecommand \@ifxundefined [1]{%
 \@ifx{#1\undefined}
}%
\providecommand \@ifnum [1]{%
 \ifnum #1\expandafter \@firstoftwo
 \else \expandafter \@secondoftwo
 \fi
}%
\providecommand \@ifx [1]{%
 \ifx #1\expandafter \@firstoftwo
 \else \expandafter \@secondoftwo
 \fi
}%
\providecommand \natexlab [1]{#1}%
\providecommand \enquote  [1]{``#1''}%
\providecommand \bibnamefont  [1]{#1}%
\providecommand \bibfnamefont [1]{#1}%
\providecommand \citenamefont [1]{#1}%
\providecommand \href@noop [0]{\@secondoftwo}%
\providecommand \href [0]{\begingroup \@sanitize@url \@href}%
\providecommand \@href[1]{\@@startlink{#1}\@@href}%
\providecommand \@@href[1]{\endgroup#1\@@endlink}%
\providecommand \@sanitize@url [0]{\catcode `\\12\catcode `\$12\catcode
  `\&12\catcode `\#12\catcode `\^12\catcode `\_12\catcode `\%12\relax}%
\providecommand \@@startlink[1]{}%
\providecommand \@@endlink[0]{}%
\providecommand \url  [0]{\begingroup\@sanitize@url \@url }%
\providecommand \@url [1]{\endgroup\@href {#1}{\urlprefix }}%
\providecommand \urlprefix  [0]{URL }%
\providecommand \Eprint [0]{\href }%
\providecommand \doibase [0]{http://dx.doi.org/}%
\providecommand \selectlanguage [0]{\@gobble}%
\providecommand \bibinfo  [0]{\@secondoftwo}%
\providecommand \bibfield  [0]{\@secondoftwo}%
\providecommand \translation [1]{[#1]}%
\providecommand \BibitemOpen [0]{}%
\providecommand \bibitemStop [0]{}%
\providecommand \bibitemNoStop [0]{.\EOS\space}%
\providecommand \EOS [0]{\spacefactor3000\relax}%
\providecommand \BibitemShut  [1]{\csname bibitem#1\endcsname}%
\let\auto@bib@innerbib\@empty
\bibitem [{\citenamefont {Rubtsov}\ \emph {et~al.}(2005)\citenamefont
  {Rubtsov}, \citenamefont {Savkin},\ and\ \citenamefont
  {Lichtenstein}}]{RubtsovPRB}%
  \BibitemOpen
  \bibfield  {author} {\bibinfo {author} {\bibfnamefont {A.~N.}\ \bibnamefont
  {Rubtsov}}, \bibinfo {author} {\bibfnamefont {V.~V.}\ \bibnamefont {Savkin}},
  \ and\ \bibinfo {author} {\bibfnamefont {A.~I.}\ \bibnamefont
  {Lichtenstein}},\ }\href {\doibase 10.1103/PhysRevB.72.035122} {\bibfield
  {journal} {\bibinfo  {journal} {Phys. Rev. B}\ }\textbf {\bibinfo {volume}
  {72}},\ \bibinfo {pages} {035122} (\bibinfo {year} {2005})},\ \Eprint
  {http://arxiv.org/abs/cond-mat/0411344} {arXiv:cond-mat/0411344
  [cond-mat.str-el]} \BibitemShut {NoStop}%
\bibitem [{\citenamefont {Werner}\ \emph {et~al.}(2006)\citenamefont {Werner},
  \citenamefont {Comanac}, \citenamefont {de’ Medici}, \citenamefont
  {Troyer},\ and\ \citenamefont {Millis}}]{werner-continuous-time-2006}%
  \BibitemOpen
  \bibfield  {author} {\bibinfo {author} {\bibfnamefont {P.}~\bibnamefont
  {Werner}}, \bibinfo {author} {\bibfnamefont {A.}~\bibnamefont {Comanac}},
  \bibinfo {author} {\bibfnamefont {L.}~\bibnamefont {de’ Medici}}, \bibinfo
  {author} {\bibfnamefont {M.}~\bibnamefont {Troyer}}, \ and\ \bibinfo {author}
  {\bibfnamefont {A.~J.}\ \bibnamefont {Millis}},\ }\href {\doibase
  10.1103/PhysRevLett.97.076405} {\bibfield  {journal} {\bibinfo  {journal}
  {Phys. Rev. Lett.}\ }\textbf {\bibinfo {volume} {97}},\ \bibinfo {pages}
  {076405} (\bibinfo {year} {2006})},\ \Eprint
  {http://arxiv.org/abs/cond-mat/0512727} {arXiv:cond-mat/0512727
  [cond-mat.str-el]} \BibitemShut {NoStop}%
\bibitem [{\citenamefont {Haule}(2007)}]{haule-quantum-2007}%
  \BibitemOpen
  \bibfield  {author} {\bibinfo {author} {\bibfnamefont {K.}~\bibnamefont
  {Haule}},\ }\href {\doibase 10.1103/PhysRevB.75.155113} {\bibfield  {journal}
  {\bibinfo  {journal} {Phys. Rev. B}\ }\textbf {\bibinfo {volume} {75}},\
  \bibinfo {pages} {155113} (\bibinfo {year} {2007})},\ \Eprint
  {http://arxiv.org/abs/cond-mat/0612172} {arXiv:cond-mat/0612172
  [cond-mat.str-el]} \BibitemShut {NoStop}%
\bibitem [{\citenamefont {Gull}\ \emph {et~al.}(2011)\citenamefont {Gull},
  \citenamefont {Millis}, \citenamefont {Lichtenstein}, \citenamefont
  {Rubtsov}, \citenamefont {Troyer},\ and\ \citenamefont
  {Werner}}]{gull-continuous-time-2011}%
  \BibitemOpen
  \bibfield  {author} {\bibinfo {author} {\bibfnamefont {E.}~\bibnamefont
  {Gull}}, \bibinfo {author} {\bibfnamefont {A.~J.}\ \bibnamefont {Millis}},
  \bibinfo {author} {\bibfnamefont {A.~I.}\ \bibnamefont {Lichtenstein}},
  \bibinfo {author} {\bibfnamefont {A.~N.}\ \bibnamefont {Rubtsov}}, \bibinfo
  {author} {\bibfnamefont {M.}~\bibnamefont {Troyer}}, \ and\ \bibinfo {author}
  {\bibfnamefont {P.}~\bibnamefont {Werner}},\ }\href {\doibase
  10.1103/RevModPhys.83.349} {\bibfield  {journal} {\bibinfo  {journal} {Rev.
  Mod. Phys.}\ }\textbf {\bibinfo {volume} {83}},\ \bibinfo {pages} {349}
  (\bibinfo {year} {2011})},\ \Eprint {http://arxiv.org/abs/1012.4474}
  {arXiv:1012.4474 [cond-mat.str-el]} \BibitemShut {NoStop}%
\bibitem [{\citenamefont {Shinaoka}\ \emph {et~al.}(2017)\citenamefont
  {Shinaoka}, \citenamefont {Gull},\ and\ \citenamefont
  {Werner}}]{shinaoka-continuous-time-2017}%
  \BibitemOpen
  \bibfield  {author} {\bibinfo {author} {\bibfnamefont {H.}~\bibnamefont
  {Shinaoka}}, \bibinfo {author} {\bibfnamefont {E.}~\bibnamefont {Gull}}, \
  and\ \bibinfo {author} {\bibfnamefont {P.}~\bibnamefont {Werner}},\ }\href
  {\doibase 10.1016/j.cpc.2017.01.003} {\bibfield  {journal} {\bibinfo
  {journal} {Comput. Phys. Commun.}\ }\textbf {\bibinfo {volume} {215}},\
  \bibinfo {pages} {128} (\bibinfo {year} {2017})},\ \Eprint
  {http://arxiv.org/abs/1609.09559} {arXiv:1609.09559 [cond-mat.str-el]}
  \BibitemShut {NoStop}%
\bibitem [{\citenamefont {Seth}\ \emph {et~al.}(2016)\citenamefont {Seth},
  \citenamefont {Krivenko}, \citenamefont {Ferrero},\ and\ \citenamefont
  {Parcollet}}]{triqs}%
  \BibitemOpen
  \bibfield  {author} {\bibinfo {author} {\bibfnamefont {P.}~\bibnamefont
  {Seth}}, \bibinfo {author} {\bibfnamefont {I.}~\bibnamefont {Krivenko}},
  \bibinfo {author} {\bibfnamefont {M.}~\bibnamefont {Ferrero}}, \ and\
  \bibinfo {author} {\bibfnamefont {O.}~\bibnamefont {Parcollet}},\ }\href
  {\doibase 10.1016/j.cpc.2015.10.023} {\bibfield  {journal} {\bibinfo
  {journal} {Comput. Phys. Commun.}\ }\textbf {\bibinfo {volume} {200}},\
  \bibinfo {pages} {274 } (\bibinfo {year} {2016})},\ \Eprint
  {http://arxiv.org/abs/1507.00175} {arXiv:1507.00175 [cond-mat.str-el]}
  \BibitemShut {NoStop}%
\bibitem [{\citenamefont {Haule}\ \emph {et~al.}(2010)\citenamefont {Haule},
  \citenamefont {Yee},\ and\ \citenamefont {Kim}}]{haule_code}%
  \BibitemOpen
  \bibfield  {author} {\bibinfo {author} {\bibfnamefont {K.}~\bibnamefont
  {Haule}}, \bibinfo {author} {\bibfnamefont {C.-H.}\ \bibnamefont {Yee}}, \
  and\ \bibinfo {author} {\bibfnamefont {K.}~\bibnamefont {Kim}},\ }\href
  {\doibase 10.1103/PhysRevB.81.195107} {\bibfield  {journal} {\bibinfo
  {journal} {Phys. Rev. B}\ }\textbf {\bibinfo {volume} {81}},\ \bibinfo
  {pages} {195107} (\bibinfo {year} {2010})},\ \Eprint
  {http://arxiv.org/abs/0907.0195} {arXiv:0907.0195 [cond-mat.str-el]}
  \BibitemShut {NoStop}%
\bibitem [{\citenamefont {Huang}\ \emph {et~al.}(2015)\citenamefont {Huang},
  \citenamefont {Wang}, \citenamefont {Meng}, \citenamefont {Du}, \citenamefont
  {Werner},\ and\ \citenamefont {Dai}}]{iqist}%
  \BibitemOpen
  \bibfield  {author} {\bibinfo {author} {\bibfnamefont {L.}~\bibnamefont
  {Huang}}, \bibinfo {author} {\bibfnamefont {Y.}~\bibnamefont {Wang}},
  \bibinfo {author} {\bibfnamefont {Z.~Y.}\ \bibnamefont {Meng}}, \bibinfo
  {author} {\bibfnamefont {L.}~\bibnamefont {Du}}, \bibinfo {author}
  {\bibfnamefont {P.}~\bibnamefont {Werner}}, \ and\ \bibinfo {author}
  {\bibfnamefont {X.}~\bibnamefont {Dai}},\ }\href {\doibase
  10.1016/j.cpc.2015.04.020} {\bibfield  {journal} {\bibinfo  {journal}
  {Comput. Phys. Commun.}\ }\textbf {\bibinfo {volume} {195}},\ \bibinfo
  {pages} {140 } (\bibinfo {year} {2015})},\ \Eprint
  {http://arxiv.org/abs/1409.7573} {arXiv:1409.7573 [cond-mat.str-el]}
  \BibitemShut {NoStop}%
\bibitem [{\citenamefont {Wallerberger}\ \emph {et~al.}(2019)\citenamefont
  {Wallerberger}, \citenamefont {Hausoel}, \citenamefont {Gunacker},
  \citenamefont {Kowalski}, \citenamefont {Parragh}, \citenamefont {Goth},
  \citenamefont {Held},\ and\ \citenamefont {Sangiovanni}}]{w2dynamics}%
  \BibitemOpen
  \bibfield  {author} {\bibinfo {author} {\bibfnamefont {M.}~\bibnamefont
  {Wallerberger}}, \bibinfo {author} {\bibfnamefont {A.}~\bibnamefont
  {Hausoel}}, \bibinfo {author} {\bibfnamefont {P.}~\bibnamefont {Gunacker}},
  \bibinfo {author} {\bibfnamefont {A.}~\bibnamefont {Kowalski}}, \bibinfo
  {author} {\bibfnamefont {N.}~\bibnamefont {Parragh}}, \bibinfo {author}
  {\bibfnamefont {F.}~\bibnamefont {Goth}}, \bibinfo {author} {\bibfnamefont
  {K.}~\bibnamefont {Held}}, \ and\ \bibinfo {author} {\bibfnamefont
  {G.}~\bibnamefont {Sangiovanni}},\ }\href {\doibase
  10.1016/j.cpc.2018.09.007} {\bibfield  {journal} {\bibinfo  {journal}
  {Comput. Phys. Commun.}\ }\textbf {\bibinfo {volume} {235}},\ \bibinfo
  {pages} {388 } (\bibinfo {year} {2019})},\ \Eprint
  {http://arxiv.org/abs/1801.10209} {arXiv:1801.10209 [cond-mat.str-el]}
  \BibitemShut {NoStop}%
\bibitem [{\citenamefont {Werner}\ and\ \citenamefont
  {Millis}(2006)}]{wernerPRB74}%
  \BibitemOpen
  \bibfield  {author} {\bibinfo {author} {\bibfnamefont {P.}~\bibnamefont
  {Werner}}\ and\ \bibinfo {author} {\bibfnamefont {A.~J.}\ \bibnamefont
  {Millis}},\ }\href {\doibase 10.1103/PhysRevB.74.155107} {\bibfield
  {journal} {\bibinfo  {journal} {Phys. Rev. B}\ }\textbf {\bibinfo {volume}
  {74}},\ \bibinfo {pages} {155107} (\bibinfo {year} {2006})},\ \Eprint
  {http://arxiv.org/abs/cond-mat/0607136} {arXiv:cond-mat/0607136
  [cond-mat.str-el]} \BibitemShut {NoStop}%
\bibitem [{\citenamefont {Parragh}\ \emph {et~al.}(2012)\citenamefont
  {Parragh}, \citenamefont {Toschi}, \citenamefont {Held},\ and\ \citenamefont
  {Sangiovanni}}]{parragh-conserved-2012}%
  \BibitemOpen
  \bibfield  {author} {\bibinfo {author} {\bibfnamefont {N.}~\bibnamefont
  {Parragh}}, \bibinfo {author} {\bibfnamefont {A.}~\bibnamefont {Toschi}},
  \bibinfo {author} {\bibfnamefont {K.}~\bibnamefont {Held}}, \ and\ \bibinfo
  {author} {\bibfnamefont {G.}~\bibnamefont {Sangiovanni}},\ }\href {\doibase
  10.1103/PhysRevB.86.155158} {\bibfield  {journal} {\bibinfo  {journal} {Phys.
  Rev. B}\ }\textbf {\bibinfo {volume} {86}},\ \bibinfo {pages} {155158}
  (\bibinfo {year} {2012})},\ \Eprint {http://arxiv.org/abs/1209.0915}
  {arXiv:1209.0915 [cond-mat.str-el]} \BibitemShut {NoStop}%
\bibitem [{\citenamefont {S\'emon}\ \emph {et~al.}(2014)\citenamefont
  {S\'emon}, \citenamefont {Yee}, \citenamefont {Haule},\ and\ \citenamefont
  {Tremblay}}]{haule-lazy-trace}%
  \BibitemOpen
  \bibfield  {author} {\bibinfo {author} {\bibfnamefont {P.}~\bibnamefont
  {S\'emon}}, \bibinfo {author} {\bibfnamefont {C.-H.}\ \bibnamefont {Yee}},
  \bibinfo {author} {\bibfnamefont {K.}~\bibnamefont {Haule}}, \ and\ \bibinfo
  {author} {\bibfnamefont {A.-M.~S.}\ \bibnamefont {Tremblay}},\ }\href
  {\doibase 10.1103/PhysRevB.90.075149} {\bibfield  {journal} {\bibinfo
  {journal} {Phys. Rev. B}\ }\textbf {\bibinfo {volume} {90}},\ \bibinfo
  {pages} {075149} (\bibinfo {year} {2014})},\ \Eprint
  {http://arxiv.org/abs/1403.7214} {arXiv:1403.7214 [cond-mat.str-el]}
  \BibitemShut {NoStop}%
\bibitem [{\citenamefont {L\"auchli}\ and\ \citenamefont
  {Werner}(2009)}]{werner-krylov-2009}%
  \BibitemOpen
  \bibfield  {author} {\bibinfo {author} {\bibfnamefont {A.~M.}\ \bibnamefont
  {L\"auchli}}\ and\ \bibinfo {author} {\bibfnamefont {P.}~\bibnamefont
  {Werner}},\ }\href {\doibase 10.1103/PhysRevB.80.235117} {\bibfield
  {journal} {\bibinfo  {journal} {Phys. Rev. B}\ }\textbf {\bibinfo {volume}
  {80}},\ \bibinfo {pages} {235117} (\bibinfo {year} {2009})},\ \Eprint
  {http://arxiv.org/abs/0908.0681} {arXiv:0908.0681 [cond-mat.str-el]}
  \BibitemShut {NoStop}%
\bibitem [{\citenamefont {Shinaoka}\ \emph {et~al.}(2014)\citenamefont
  {Shinaoka}, \citenamefont {Dolfi}, \citenamefont {Troyer},\ and\
  \citenamefont {Werner}}]{shinaoka-hybridization-2014}%
  \BibitemOpen
  \bibfield  {author} {\bibinfo {author} {\bibfnamefont {H.}~\bibnamefont
  {Shinaoka}}, \bibinfo {author} {\bibfnamefont {M.}~\bibnamefont {Dolfi}},
  \bibinfo {author} {\bibfnamefont {M.}~\bibnamefont {Troyer}}, \ and\ \bibinfo
  {author} {\bibfnamefont {P.}~\bibnamefont {Werner}},\ }\href {\doibase
  10.1088/1742-5468/2014/06/P06012} {\bibfield  {journal} {\bibinfo  {journal}
  {J. Stat. Mech. Theory Exp.}\ }\textbf {\bibinfo {volume} {2014}},\ \bibinfo
  {pages} {P06012} (\bibinfo {year} {2014})},\ \Eprint
  {http://arxiv.org/abs/1404.1259} {arXiv:1404.1259 [cond-mat.str-el]}
  \BibitemShut {NoStop}%
\bibitem [{\citenamefont {Augustinsk{\'{y}}}\ and\ \citenamefont
  {Kune{\v{s}}}(2013)}]{Augustinsky2013}%
  \BibitemOpen
  \bibfield  {author} {\bibinfo {author} {\bibfnamefont {P.}~\bibnamefont
  {Augustinsk{\'{y}}}}\ and\ \bibinfo {author} {\bibfnamefont {J.}~\bibnamefont
  {Kune{\v{s}}}},\ }\href {\doibase 10.1016/j.cpc.2013.04.005} {\bibfield
  {journal} {\bibinfo  {journal} {Comput. Phys. Commun.}\ }\textbf {\bibinfo
  {volume} {184}},\ \bibinfo {pages} {2119} (\bibinfo {year} {2013})},\ \Eprint
  {http://arxiv.org/abs/1302.4594} {arXiv:1302.4594} \BibitemShut {NoStop}%
\bibitem [{\citenamefont {Wallerberger}(2016)}]{wallerberger-w2dynamics-2016}%
  \BibitemOpen
  \bibfield  {author} {\bibinfo {author} {\bibfnamefont {M.}~\bibnamefont
  {Wallerberger}},\ }\emph {\bibinfo {title} {w2dynamics: continuous time
  quantum {Monte} {Carlo} calculations of one- and two-particle propagators}},\
  \href {https://resolver.obvsg.at/urn:nbn:at:at-ubtuw:1-3537} {Ph.D. thesis},\
  \bibinfo  {school} {Technische Universit{\"a}t Wien}, \bibinfo {address}
  {Wien} (\bibinfo {year} {2016})\BibitemShut {NoStop}%
\bibitem [{Note1()}]{Note1}%
  \BibitemOpen
  \bibinfo {note} {One may somewhat reduce the residual advantage of superstate
  sampling visible at large $\beta $ in Fig.~\ref {fig:speedup} upon optimizing
  the quantum number checking or using more sophisticated schemes, such as
  those proposed in Ref.~\cite {haule-lazy-trace}}\BibitemShut {NoStop}%
\bibitem [{Note3({\natexlab{a}})}]{CNoteModel}%
  \BibitemOpen
  \bibinfo {note} {We used a five-orbital Anderson impurity model with a local
  Slater--Kanamori interaction, parameterized by $U \approx 5.03$, $J \approx
  0.64$, and $U^\prime = U - 2J$. The temperature was $\beta = 30$ where not
  stated otherwise and the chemical potential was chosen to reach a mean
  filling of $\langle N\rangle = 8$ in the interacting model. The hybridization
  function was obtained from DFT for a Cobalt impurity on Cu(001) as described
  in Ref.~\onlinecite{bahlke-interplay-2018}.}\BibitemShut {Stop}%
\bibitem [{\citenamefont {Bahlke}\ \emph {et~al.}(2018)\citenamefont {Bahlke},
  \citenamefont {Karolak},\ and\ \citenamefont
  {Herrmann}}]{bahlke-interplay-2018}%
  \BibitemOpen
  \bibfield  {author} {\bibinfo {author} {\bibfnamefont {M.~P.}\ \bibnamefont
  {Bahlke}}, \bibinfo {author} {\bibfnamefont {M.}~\bibnamefont {Karolak}}, \
  and\ \bibinfo {author} {\bibfnamefont {C.}~\bibnamefont {Herrmann}},\ }\href
  {\doibase 10.1103/PhysRevB.97.035119} {\bibfield  {journal} {\bibinfo
  {journal} {Phys. Rev. B}\ }\textbf {\bibinfo {volume} {97}},\ \bibinfo
  {pages} {035119} (\bibinfo {year} {2018})},\ \Eprint
  {http://arxiv.org/abs/1710.07349} {arXiv:1710.07349 [cond-mat.str-el]}
  \BibitemShut {NoStop}%
\bibitem [{Note2({\natexlab{a}})}]{CNote1}%
  \BibitemOpen
  \bibinfo {note} {Note that the number of states per superstate varies and the
  axis range of 10 was sufficient to capture the maximum number of contributing
  states over the entire course of the simulation due to many good quantum
  numbers.}\BibitemShut {Stop}%
\bibitem [{Note2({\natexlab{b}})}]{Note2}%
  \BibitemOpen
  \bibinfo {note} {Specifically, a matrix-vector implementation of CT-HYB with
  time evolution in eigenbasis. Quantum number checking (cf.\ Sec.~\ref
  {s_cthyb}) is used to avoid the calculation of zero contributions to the
  local weight using a partitioning of the Hilbert space into superstates using
  conserved quantities for Kanamori-like interaction\cite
  {parragh-conserved-2012} and additionally automatic partitioning\cite
  {haule-quantum-2007,triqs} for full Coulomb interaction. We do not use outer
  truncation of the local trace, sliding-window-style local updates, tree
  algorithms, or other optimizations even if mentioned in the introduction
  unless explicitly stated.}\BibitemShut {Stop}%
\bibitem [{Note3({\natexlab{b}})}]{Note3}%
  \BibitemOpen
  \bibinfo {note} {We used about the same amount of CPU time for all sampling
  methods for this example, so due to the worse performance of the conventional
  implementation its error is considerably larger. While the error in this
  example is thus sufficiently large to allow other conclusions about the
  relative sign of the methods in some temperature ranges, data from many other
  calculations we did with all methods not specifically for the purpose of this
  article strongly indicate equal signs for conventional and superstate
  sampling and a sign closer to zero (with model-dependent extent) for state
  sampling.}\BibitemShut {Stop}%
\bibitem [{Note4()}]{Note4}%
  \BibitemOpen
  \bibinfo {note} {In this work we consider the computational cost of
  quantum-number checking in \protect \emph {w2dynamics} \cite {w2dynamics}
  with decreasing temperature. The magnitude of this speed up might of course
  be lower in other implementations.}\BibitemShut {Stop}%
\bibitem [{\citenamefont {Kowalski}(2017)}]{kowalski-improved-2017}%
  \BibitemOpen
  \bibfield  {author} {\bibinfo {author} {\bibfnamefont {A.}~\bibnamefont
  {Kowalski}},\ }\emph {\bibinfo {title} {Improved sampling in continuous-time
  quantum Monte Carlo algorithms for fermions}},\ \href@noop {} {Master's
  thesis},\ \bibinfo  {school} {Julius-Maximilians-Universit\"at W\"urzburg},
  \bibinfo {address} {W\"urzburg} (\bibinfo {year} {2017})\BibitemShut
  {NoStop}%
\bibitem [{\citenamefont {Werner}\ \emph {et~al.}(2008)\citenamefont {Werner},
  \citenamefont {Gull}, \citenamefont {Troyer},\ and\ \citenamefont
  {Millis}}]{werner-spin-2008}%
  \BibitemOpen
  \bibfield  {author} {\bibinfo {author} {\bibfnamefont {P.}~\bibnamefont
  {Werner}}, \bibinfo {author} {\bibfnamefont {E.}~\bibnamefont {Gull}},
  \bibinfo {author} {\bibfnamefont {M.}~\bibnamefont {Troyer}}, \ and\ \bibinfo
  {author} {\bibfnamefont {A.~J.}\ \bibnamefont {Millis}},\ }\href {\doibase
  10.1103/PhysRevLett.101.166405} {\bibfield  {journal} {\bibinfo  {journal}
  {Phys. Rev. Lett.}\ }\textbf {\bibinfo {volume} {101}},\ \bibinfo {pages}
  {166405} (\bibinfo {year} {2008})},\ \Eprint {http://arxiv.org/abs/0806.2621}
  {arXiv:0806.2621 [cond-mat.str-el]} \BibitemShut {NoStop}%
\bibitem [{\citenamefont {{de' Medici}}\ \emph {et~al.}(2011)\citenamefont
  {{de' Medici}}, \citenamefont {Mravlje},\ and\ \citenamefont
  {Georges}}]{de-medici-janus-faced-2011}%
  \BibitemOpen
  \bibfield  {author} {\bibinfo {author} {\bibfnamefont {L.}~\bibnamefont {{de'
  Medici}}}, \bibinfo {author} {\bibfnamefont {J.}~\bibnamefont {Mravlje}}, \
  and\ \bibinfo {author} {\bibfnamefont {A.}~\bibnamefont {Georges}},\ }\href
  {\doibase 10.1103/PhysRevLett.107.256401} {\bibfield  {journal} {\bibinfo
  {journal} {Phys. Rev. Lett.}\ }\textbf {\bibinfo {volume} {107}},\ \bibinfo
  {pages} {256401} (\bibinfo {year} {2011})},\ \Eprint
  {http://arxiv.org/abs/1106.0815} {arXiv:1106.0815 [cond-mat.str-el]}
  \BibitemShut {NoStop}%
\bibitem [{\citenamefont {De~Leo}(2004)}]{de-leo-non-fermi-2004}%
  \BibitemOpen
  \bibfield  {author} {\bibinfo {author} {\bibfnamefont {L.}~\bibnamefont
  {De~Leo}},\ }\emph {\bibinfo {title} {Non-{Fermi} liquid behavior in
  multi-orbital {Anderson} impurity models and possible relevance for strongly
  correlated lattice models}},\ \href
  {https://iris.sissa.it/handle/20.500.11767/4016} {Ph.D. thesis},\ \bibinfo
  {school} {SISSA} (\bibinfo {year} {2004})\BibitemShut {NoStop}%
\bibitem [{\citenamefont {De~Leo}\ and\ \citenamefont
  {Fabrizio}(2004)}]{leo-spectral-2004}%
  \BibitemOpen
  \bibfield  {author} {\bibinfo {author} {\bibfnamefont {L.}~\bibnamefont
  {De~Leo}}\ and\ \bibinfo {author} {\bibfnamefont {M.}~\bibnamefont
  {Fabrizio}},\ }\href {\doibase 10.1103/PhysRevB.69.245114} {\bibfield
  {journal} {\bibinfo  {journal} {Phys. Rev. B}\ }\textbf {\bibinfo {volume}
  {69}},\ \bibinfo {pages} {245114} (\bibinfo {year} {2004})},\ \Eprint
  {http://arxiv.org/abs/cond-mat/0402121} {arXiv:cond-mat/0402121
  [cond-mat.str-el]} \BibitemShut {NoStop}%
\bibitem [{\citenamefont {Yin}\ \emph {et~al.}(2012)\citenamefont {Yin},
  \citenamefont {Haule},\ and\ \citenamefont {Kotliar}}]{haulePRB86}%
  \BibitemOpen
  \bibfield  {author} {\bibinfo {author} {\bibfnamefont {Z.~P.}\ \bibnamefont
  {Yin}}, \bibinfo {author} {\bibfnamefont {K.}~\bibnamefont {Haule}}, \ and\
  \bibinfo {author} {\bibfnamefont {G.}~\bibnamefont {Kotliar}},\ }\href
  {\doibase 10.1103/PhysRevB.86.195141} {\bibfield  {journal} {\bibinfo
  {journal} {Phys. Rev. B}\ }\textbf {\bibinfo {volume} {86}},\ \bibinfo
  {pages} {195141} (\bibinfo {year} {2012})},\ \Eprint
  {http://arxiv.org/abs/1206.0801} {arXiv:1206.0801 [cond-mat.str-el]}
  \BibitemShut {NoStop}%
\bibitem [{\citenamefont {Georges}\ \emph {et~al.}(2013)\citenamefont
  {Georges}, \citenamefont {Medici},\ and\ \citenamefont
  {Mravlje}}]{georges-strong-2013}%
  \BibitemOpen
  \bibfield  {author} {\bibinfo {author} {\bibfnamefont {A.}~\bibnamefont
  {Georges}}, \bibinfo {author} {\bibfnamefont {L.~d.}\ \bibnamefont {Medici}},
  \ and\ \bibinfo {author} {\bibfnamefont {J.}~\bibnamefont {Mravlje}},\ }\href
  {\doibase 10.1146/annurev-conmatphys-020911-125045} {\bibfield  {journal}
  {\bibinfo  {journal} {Annu. Rev. Condens. Matter Phys.}\ }\textbf {\bibinfo
  {volume} {4}},\ \bibinfo {pages} {137} (\bibinfo {year} {2013})},\ \Eprint
  {http://arxiv.org/abs/1207.3033} {arXiv:1207.3033 [cond-mat.str-el]}
  \BibitemShut {NoStop}%
\bibitem [{\citenamefont {Stadler}\ \emph {et~al.}(2018)\citenamefont
  {Stadler}, \citenamefont {Kotliar}, \citenamefont {Weichselbaum},\ and\
  \citenamefont {von Delft}}]{stadler-hundness-2018}%
  \BibitemOpen
  \bibfield  {author} {\bibinfo {author} {\bibfnamefont {K.}~\bibnamefont
  {Stadler}}, \bibinfo {author} {\bibfnamefont {G.}~\bibnamefont {Kotliar}},
  \bibinfo {author} {\bibfnamefont {A.}~\bibnamefont {Weichselbaum}}, \ and\
  \bibinfo {author} {\bibfnamefont {J.}~\bibnamefont {von Delft}},\ }\href
  {\doibase 10.1016/j.aop.2018.10.017} {\bibfield  {journal} {\bibinfo
  {journal} {Annals of Physics}\ } (\bibinfo {year} {2018}),\
  10.1016/j.aop.2018.10.017},\ \Eprint {http://arxiv.org/abs/1808.09936}
  {arXiv:1808.09936 [cond-mat.str-el]} \BibitemShut {NoStop}%
\bibitem [{\citenamefont {Kim}\ \emph {et~al.}(2017)\citenamefont {Kim},
  \citenamefont {Jeschke}, \citenamefont {Werner},\ and\ \citenamefont
  {Valent{\'\i}}}]{kim-j-2017}%
  \BibitemOpen
  \bibfield  {author} {\bibinfo {author} {\bibfnamefont {A.~J.}\ \bibnamefont
  {Kim}}, \bibinfo {author} {\bibfnamefont {H.~O.}\ \bibnamefont {Jeschke}},
  \bibinfo {author} {\bibfnamefont {P.}~\bibnamefont {Werner}}, \ and\ \bibinfo
  {author} {\bibfnamefont {R.}~\bibnamefont {Valent{\'\i}}},\ }\href {\doibase
  10.1103/PhysRevLett.118.086401} {\bibfield  {journal} {\bibinfo  {journal}
  {Phys. Rev. Lett.}\ }\textbf {\bibinfo {volume} {118}},\ \bibinfo {pages}
  {086401} (\bibinfo {year} {2017})},\ \Eprint
  {http://arxiv.org/abs/1607.05196} {arXiv:1607.05196 [cond-mat.str-el]}
  \BibitemShut {NoStop}%
\bibitem [{\citenamefont {Chubukov}\ and\ \citenamefont
  {Maslov}(2012)}]{chubukov-first-matsubara-frequency-2012}%
  \BibitemOpen
  \bibfield  {author} {\bibinfo {author} {\bibfnamefont {A.~V.}\ \bibnamefont
  {Chubukov}}\ and\ \bibinfo {author} {\bibfnamefont {D.~L.}\ \bibnamefont
  {Maslov}},\ }\href {\doibase 10.1103/PhysRevB.86.155136} {\bibfield
  {journal} {\bibinfo  {journal} {Phys. Rev. B}\ }\textbf {\bibinfo {volume}
  {86}},\ \bibinfo {pages} {155136} (\bibinfo {year} {2012})},\ \Eprint
  {http://arxiv.org/abs/1208.3483} {arXiv:1208.3483 [cond-mat.str-el]}
  \BibitemShut {NoStop}%
\bibitem [{\citenamefont {Pourovskii}\ \emph {et~al.}(2017)\citenamefont
  {Pourovskii}, \citenamefont {Mravlje}, \citenamefont {Georges}, \citenamefont
  {Simak},\ and\ \citenamefont {Abrikosov}}]{pourovskii-electronelectron-2017}%
  \BibitemOpen
  \bibfield  {author} {\bibinfo {author} {\bibfnamefont {L.~V.}\ \bibnamefont
  {Pourovskii}}, \bibinfo {author} {\bibfnamefont {J.}~\bibnamefont {Mravlje}},
  \bibinfo {author} {\bibfnamefont {A.}~\bibnamefont {Georges}}, \bibinfo
  {author} {\bibfnamefont {S.~I.}\ \bibnamefont {Simak}}, \ and\ \bibinfo
  {author} {\bibfnamefont {I.~A.}\ \bibnamefont {Abrikosov}},\ }\href {\doibase
  10.1088/1367-2630/aa76c9} {\bibfield  {journal} {\bibinfo  {journal} {New
  Journal of Physics}\ }\textbf {\bibinfo {volume} {19}},\ \bibinfo {pages}
  {073022} (\bibinfo {year} {2017})},\ \Eprint
  {http://arxiv.org/abs/1603.02287} {arXiv:1603.02287 [cond-mat.str-el]}
  \BibitemShut {NoStop}%
\bibitem [{\citenamefont {Hausoel}\ \emph {et~al.}(2017)\citenamefont
  {Hausoel}, \citenamefont {Karolak}, \citenamefont {\c{S}a\c{s}{\i}o\u{g}lu},
  \citenamefont {Lichtenstein}, \citenamefont {Held}, \citenamefont {Katanin},
  \citenamefont {Toschi},\ and\ \citenamefont
  {Sangiovanni}}]{hausoel-local-2017}%
  \BibitemOpen
  \bibfield  {author} {\bibinfo {author} {\bibfnamefont {A.}~\bibnamefont
  {Hausoel}}, \bibinfo {author} {\bibfnamefont {M.}~\bibnamefont {Karolak}},
  \bibinfo {author} {\bibfnamefont {E.}~\bibnamefont
  {\c{S}a\c{s}{\i}o\u{g}lu}}, \bibinfo {author} {\bibfnamefont
  {A.}~\bibnamefont {Lichtenstein}}, \bibinfo {author} {\bibfnamefont
  {K.}~\bibnamefont {Held}}, \bibinfo {author} {\bibfnamefont {A.}~\bibnamefont
  {Katanin}}, \bibinfo {author} {\bibfnamefont {A.}~\bibnamefont {Toschi}}, \
  and\ \bibinfo {author} {\bibfnamefont {G.}~\bibnamefont {Sangiovanni}},\
  }\href {\doibase 10.1038/ncomms16062} {\bibfield  {journal} {\bibinfo
  {journal} {Nature Communications}\ }\textbf {\bibinfo {volume} {8}},\
  \bibinfo {pages} {16062} (\bibinfo {year} {2017})},\ \Eprint
  {http://arxiv.org/abs/1707.03789} {arXiv:1707.03789 [cond-mat.str-el]}
  \BibitemShut {NoStop}%
\bibitem [{\citenamefont {Wilson}(1975)}]{wilsonRMP}%
  \BibitemOpen
  \bibfield  {author} {\bibinfo {author} {\bibfnamefont {K.~G.}\ \bibnamefont
  {Wilson}},\ }\href {\doibase 10.1103/RevModPhys.47.773} {\bibfield  {journal}
  {\bibinfo  {journal} {Rev. Mod. Phys.}\ }\textbf {\bibinfo {volume} {47}},\
  \bibinfo {pages} {773} (\bibinfo {year} {1975})}\BibitemShut {NoStop}%
\bibitem [{\citenamefont {Stadler}\ \emph {et~al.}(2015)\citenamefont
  {Stadler}, \citenamefont {Yin}, \citenamefont {von Delft}, \citenamefont
  {Kotliar},\ and\ \citenamefont {Weichselbaum}}]{stadlerPRL}%
  \BibitemOpen
  \bibfield  {author} {\bibinfo {author} {\bibfnamefont {K.~M.}\ \bibnamefont
  {Stadler}}, \bibinfo {author} {\bibfnamefont {Z.~P.}\ \bibnamefont {Yin}},
  \bibinfo {author} {\bibfnamefont {J.}~\bibnamefont {von Delft}}, \bibinfo
  {author} {\bibfnamefont {G.}~\bibnamefont {Kotliar}}, \ and\ \bibinfo
  {author} {\bibfnamefont {A.}~\bibnamefont {Weichselbaum}},\ }\href {\doibase
  10.1103/PhysRevLett.115.136401} {\bibfield  {journal} {\bibinfo  {journal}
  {Phys. Rev. Lett.}\ }\textbf {\bibinfo {volume} {115}},\ \bibinfo {pages}
  {136401} (\bibinfo {year} {2015})},\ \Eprint
  {http://arxiv.org/abs/1503.06467} {arXiv:1503.06467 [cond-mat.str-el]}
  \BibitemShut {NoStop}%
\bibitem [{\citenamefont {Horvat}\ \emph {et~al.}(2016)\citenamefont {Horvat},
  \citenamefont {\ifmmode~\check{Z}\else \v{Z}\fi{}itko},\ and\ \citenamefont
  {Mravlje}}]{horvatPRB}%
  \BibitemOpen
  \bibfield  {author} {\bibinfo {author} {\bibfnamefont {A.}~\bibnamefont
  {Horvat}}, \bibinfo {author} {\bibfnamefont {R.}~\bibnamefont
  {\ifmmode~\check{Z}\else \v{Z}\fi{}itko}}, \ and\ \bibinfo {author}
  {\bibfnamefont {J.}~\bibnamefont {Mravlje}},\ }\href {\doibase
  10.1103/PhysRevB.94.165140} {\bibfield  {journal} {\bibinfo  {journal} {Phys.
  Rev. B}\ }\textbf {\bibinfo {volume} {94}},\ \bibinfo {pages} {165140}
  (\bibinfo {year} {2016})},\ \Eprint {http://arxiv.org/abs/1606.07654}
  {arXiv:1606.07654 [cond-mat.str-el]} \BibitemShut {NoStop}%
\bibitem [{\citenamefont {{de' Medici}}(2017)}]{lucaLecture}%
  \BibitemOpen
  \bibfield  {author} {\bibinfo {author} {\bibfnamefont {L.}~\bibnamefont {{de'
  Medici}}},\ }\enquote {\bibinfo {title} {Hund's metals explained},}\ in\
  \href {https://juser.fz-juelich.de/record/837488} {\emph {\bibinfo
  {booktitle} {{T}he {P}hysics of {C}orrelated {I}nsulators, {M}etals, and
  {S}uperconductors}}},\ \bibinfo {series} {Schriften des Forschungszentrums
  Jülich Reihe Modeling and Simulation}, Vol.~\bibinfo {volume} {7},\ \bibinfo
  {editor} {edited by\ \bibinfo {editor} {\bibfnamefont {E.}~\bibnamefont
  {Pavarini}}, \bibinfo {editor} {\bibfnamefont {E.}~\bibnamefont {Koch}},
  \bibinfo {editor} {\bibfnamefont {R.}~\bibnamefont {Scalettar}}, \ and\
  \bibinfo {editor} {\bibfnamefont {R.}~\bibnamefont {Martin}}}\ (\bibinfo
  {publisher} {Forschungszentrum Jülich GmbH Zentralbibliothek, Verlag},\
  \bibinfo {address} {Jülich},\ \bibinfo {year} {2017})\ Chap.~\bibinfo
  {chapter} {14}\BibitemShut {NoStop}%
\end{thebibliography}%

\end{document}